\documentclass{eas}
\usepackage{graphicx}
\usepackage{amssymb}

\newcommand\lta{\mathrel{\hbox{\raise 0.6 ex \hbox{$<$}\kern
                   -1.8 ex\lower .5 ex\hbox{$\sim$}}}}
\newcommand\gta{\mathrel{\hbox{\raise 0.6 ex \hbox{$>$}\kern
                   -1.7 ex\lower .5 ex\hbox{$\sim$}}}}
\newcommand{\beqa}{\begin{eqnarray*}}
\newcommand{\eeqa}{\end{eqnarray*}}
\newcommand{\beqan}{\begin{eqnarray}}
\newcommand{\eeqan}{\end{eqnarray}}
\newcommand{\beq}{\begin{equation}}
\newcommand{\eeq}{\end{equation}}
\newcommand{\drr}{\frac{\partial}{\partial r}}
\newcommand{\dtt}{\frac{\diff}{\diff t}}
\newcommand{\dr}[1]{\frac{\partial  #1}{\partial r}}
\newcommand{\dt}[1]{\frac{\partial  #1}{\partial t}}
\newcommand{\lp}{ \left(}
\newcommand{\rp}{ \right)}
\newcommand{\lc}{ \left[}
\newcommand{\rc}{ \right]}
\newcommand{\diff}{{\rm d}}
\newcommand{\eg}{{\em e.g.\,}}
\renewcommand{\cf}{{\em cf.\,}}

\begin{document}

\title{Transport processes in stars: diffusion, rotation, magnetic
fields and internal waves}

\runningtitle{S.\,Talon: Transport processes in stars}
\author{Suzanne Talon}
\address{D\'epartement de physique, Universit\'e de Montr\'eal, C.P. 6128,
succursale centre-ville, Montr\'eal, Qu\'ebec, H3C 3J8}

\begin{abstract}
In this paper, I explore various transport processes that have 
a large impact of the distribution of elements inside stars
and thus, on stellar evolution. A heuristic description of the physics behind
equations is provided, and key references are given. Finally, for
each process, I will briefly review (some) important results as well as discuss
directions for future work. 
\end{abstract}

\maketitle

\section{Atomic diffusion \label{sec:atomic}}

\subsection{Introduction}

Atomic diffusion refers to atomic scale 
physical processes that cause the stratification
of various species and the smoothing of chemical discontinuities. 
These processes occur when multicomponent plasmas are subject to 
physical conditions that
vary within the medium (such as pressure, temperature or density) or
are subject to external forces that vary from species to species
and are related to the Brownian motions of particles.
Stellar interiors are a good laboratory for the study of these processes.

As atomic diffusion is not the main subject of this paper, this section
will briefly outline general ideas to keep in mind when diffusion\footnote{Let
us note that atomic diffusion is generally referred to simply as {\em
diffusion}.} is mentioned. Calculations are presented for a
small concentration of particles $i$
diffusing in protons $p$. This is called the {\em trace element approximation}.
For a more complete description of diffusion in a general context
see Burgers~(1969) or Chapman \& Cowling~(1970).

\subsection{Diffusion and external forces}

We first consider a gas in which physical parameters are constant and some external
force ${\cal F}_{{\rm ext}_i}$ is exerted on particles $i$. 
The total force on $i$ is then
\beq
{\cal F}_i =
-k_BT\dr{\ln P_i}+ {\cal F}_{\rm ext_i} \label{dif:F}
\eeq
where $P_i$ represents the partial pressure on $i$, and is related to the
total pressure $P$ and the concentration $c_i$ by
\beq
c_i = \frac{P_i}{P_p+P_i}~~~{\rm with}~P_i \ll P_p.
\eeq
Equilibrium is obtained when ${\cal F}_i =0$. Since diffusion is a slow process
that should bring composition
gradients to their equilibrium values, it is reasonable to
assume that the diffusion velocity $V_{ip}$ 
of $i$ with respect to protons is proportional to
the difference between actual and equilibrium values 
(Eddington~1930) and thus
\beq
V_{ip} = D_{\rm at} \lc -\dr{\ln c_i} + \frac{{\cal F}_{{\rm ext}_i}}{k_BT} \rc
\eeq
and the change of $c_i$ during this evolution is given by\footnote{Here,
modifications of the structure are neglected.}
\beq
\dt{c_i} = -\drr \lp V_{ip} c_i \rp.
\eeq
The atomic diffusion coefficient is
$D_{\rm at} = \frac{1}{3} \Lambda v_T$ with $\Lambda$ the particle mean free path
and $v_T$ the thermal velocity (see \eg Aller \& Chapman~1960). 
In the absence of an external force, this
equation shows the tendency of microscopic motions to eliminate abundance
gradients.

\subsection{Gravitational settling and radiative forces}

Let us now look at two specific external forces encountered in stellar
interiors. First, we have gravity. When the mean molecular mass of the stellar
plasma $\mu$
is different from the mass of $i$ ($m_i$), the differential force on $i$
is
\beq
{\cal F}_{{\rm ext}_i} = \mu g - m_i g =  (\mu - m_i)g
\eeq
where we used again the perfect gas law, and assumed hydrostatic equilibrium.
$g$ is the local gravitational acceleration.
In stars, this equation implies that elements heavier than hydrogen must
``fall'' towards the star's center, and this is called {\em gravitational
settling}.

\begin{figure}[t]
\begin{center}
\includegraphics[width=7.5cm]{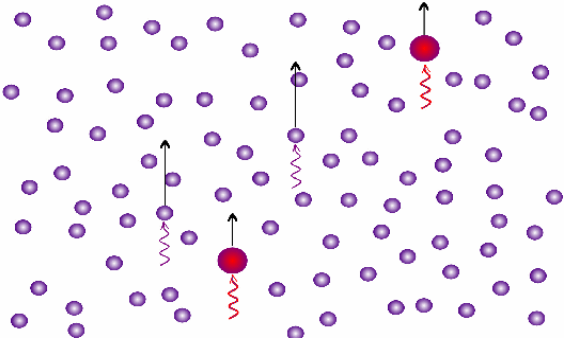}
\end{center}
\caption{The absorption of photons of different energies by various
atomic species causes an acceleration that opposes gravity. When
absorption is more probable, and the number of absorbers is small, this may
produce their {\em radiative levitation}.
\label{Frad} }
\end{figure}

The other force that has a large impact on element stratification
is the transfer of momentum caused by photon absorption. 
In the region where radiative transfer can be described using the
diffusion approximation, the star's thermal
structure is calculated using the {\em Rosseland mean opacity} $\kappa_R$
defined as
\beq
\frac{1}{\kappa_R} \int_0^\infty 
\frac{\partial B(\nu,T)}{\partial T} \diff \nu =  \int_0^\infty \frac{1}{\kappa_\nu} 
\frac{\partial B(\nu,T)}{\partial T} \diff \nu
\eeq
where $B(\nu,T)$ is the Planck function and $\kappa_\nu$ the monochromatic
opacity. 
The radiative luminosity $L^{\rm rad}$ is then related to the temperature 
gradient by
\beq
L^{\rm rad}= -\frac{64 \pi \sigma_R r^2}{3} \frac{T^3}{\kappa_R \rho} 
\frac{\diff T}{\diff r}
\eeq
where $\sigma_R$ is the Stefan-Boltzmann constant.
For a given element $i$, the absorbed momentum is proportional to the
fraction $f_i$ of the total flux that is absorbed\footnote{This is in fact an
approximation. The complete treatment involves momentum redistribution
between ions and electrons (see Richer et al.~1998 for details).} by that type of particle.
We can now compute the radiative acceleration of particle $i$ as
\beq
{g_{\rm rad}}_i=\frac{1}{4\pi r^2} \frac{L^{\rm rad}}{c} \kappa_R
\frac{f_i}{X_i} 
\eeq
where 
\beq
f_i=\int_0^\infty \frac{{\kappa_\nu}_i}{\kappa_\nu} {\cal P}(u)\diff u
~~~{\rm with}~~~\int_0^\infty{\cal P}(u)\diff u=1
\eeq
where $u=h\nu/k_BT$ and 
${\cal P}(u)=\frac{15}{4\pi^4}\frac{ u^4e^u}{(e^u-1)^2}$ 
is the normalized radiation flux associated with the temperature
derivative of the Planck function
(this was derived first by Michaud et al.~1976).
The calculation of radiative forces thus 
requires detailed spectra for all species and all ionization states.

For a given trace element, when $g_{\rm rad} > g$, the element will be
``supported'' by radiation, and will move upwards toward regions
of force neutrality ($g_{\rm rad} \simeq g$); this neutrality condition may
also result from absorption saturation\footnote{As atoms migrate
upwards and $X_i$ rises, they have to
share the momentum that photons give away, potentially 
leading to a saturation of
$g_{\rm rad}$ which limits further growth of $X_i$.}. 
In the opposite
case ($g_{\rm rad} <g$), elements sink.

\subsection{Diffusion and temperature gradients}

Temperature gradients are another source of element stratification. In a binary
fluid (with all other physical parameters constant), the particles mean
free path depend on temperature through their collision cross section (roughly as
$\sigma \propto T^{-2}$ for ions, while it does not depend on $T$ for neutral
atoms) and through their thermal velocity ($v_T \propto T^{1/2}$).

\begin{figure}[t]
\begin{center}
\includegraphics[width=7.5cm]{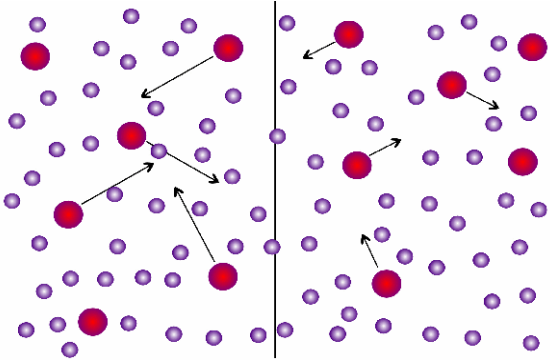}
\end{center}
\caption{When temperature is not homogeneous, the particles mean free path 
varies. In this cartoon, the probability of crossing the line 
to the left is different from the probability of
crossing it to the right if the abundance is homogeneous. The achievement of equilibrium
requires that there be a lower density of big particles on the left than on the right.
\label{D_therm} }
\end{figure}

The efficiency of this process can be estimated by considering the net flux of
ions through an equipotential (Vauclair,~1983)
\beq
n_i V_i = \frac{1}{6} \left. v_T \right|_{r+\Lambda} \left. n_i \right|_{r+\Lambda} 
\frac{\left. \sigma \right|_{r+\Lambda}}{\left. \sigma \right|_r} -
\frac{1}{6} \left. v_T \right|_{r-\Lambda} \left. n_i \right|_{r-\Lambda} 
\frac{\left. \sigma \right|_{r-\Lambda}}{\left. \sigma \right|_r} 
\eeq
where $n_i$ is the number density and assuming that on average $1/6$ of 
ions move in the direction of the equipotential. If $n_i$ is fixed, this is simply
\beq
V_i = \frac{\Lambda}{3\sigma} \dr{\lp v_T \sigma \rp}
\simeq -\frac{1}{2} D_{\rm at} \dr{\ln T} \label{vth}
\eeq
for ions. A realistic calculation would give a leading factor of $2.65 (Z_i/Z_p)^2$
instead of the $2$ given here. In more general conditions, even the sign in
Eq.~(\ref{vth}) may be different as it depends on the masses of the colliding
particles. This is discussed at length for example in Michaud~(1993).

\begin{figure}[t]
\begin{center}
\includegraphics[width=6.5cm]{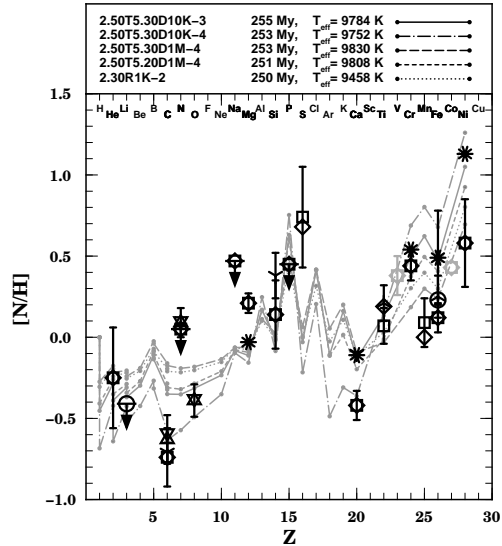}
\vspace*{-0.7cm}
\end{center}
\caption{Surface abundances in Sirius under the effect of atomic diffusion including radiative
forces and some ad-hoc turbulence (various lines) compared with observations
by various authors (symbols). (Figure 18 of Richer et al.~2000.)
\label{sirius} }
\end{figure}

\subsection{Some applications of diffusion theory \label{app:dif}}

Diffusion theory has been applied to a wide variety of stars. Its effects are expected to be
largest in stars which have only shallow surface convection zones. 
This is the case in particular of AmFm stars, 
which exhibit abundance anomalies that can be as large as one order of magnitude for certain
elements and thus, it is believed that these are only superficial. The Montreal group 
calculated stellar models including all effects of atomic diffusion, including 
radiative accelerations (Richer, Michaud
\& Turcotte,~2000). One of their results is shown on Fig.~\ref{sirius}. 
When radiative forces alone are considered, the predicted anomalies are
too large by a factor of $\sim3$. In the results shown here, a small amount of ad-hoc
turbulence has been added to obtain a good agreement with the observations.

Diffusion theory has also been applied to solar models. There, it predicts the settling of
helium below the convection zone: such settling actually improves the agreement with
helioseismology as was originally shown by Christensen-Dalsgaard, Proffitt
\& Thompson (1993). The impact of thermal diffusion on solar models
has been studied with great care
by Turcotte et al.~(1998) who showed that, for some elements, it increases the
settling velocity by as much as 50\%.

Finally, diffusion can also have significant effects on the stellar structure
when slow changes can accumulate over the stellar lifetime. In old population~II
stars, it is found to reduce the age determination by isochrones in globular
clusters by about  10\% (VandenBerg et al.~2002).

\section{Rotational mixing}

\subsection{Introduction}
Descriptions of rotational mixing have evolved a lot in recent
years. The need for extra mixing in stars is widely recognized, and
different rotational histories could well explain the variety of
stellar behaviors.

Early descriptions have relied on very simple power laws relating the
amount of mixing to the star's rotation rate (see {\em e.g.} Zahn~1983). 
In the case of 
lithium destruction in low mass stars
for example, this would lead to larger dispersion in Li
than observed (Schatzman \& Baglin~1991).

Very early on, it was recognized that such scaling laws
fail to produce an appropriate depth dependent diffusion profile and that
consistent models must be built that take into account the 
angular momentum evolution (Endal \& Sofia~1978).
To do so, one must consider all possible transport processes for
angular momentum and then analyze their effect on the transport of chemicals.

\subsection{Rotation driven instabilities \label{sec:rot}}

In this first section, we will be looking at hydrodynamical
instabilities. They are presented by considering what happens if one
displaces a fluid parcel from its original position.

\begin{figure}[t]
\begin{center}
\includegraphics[width=5.5cm]{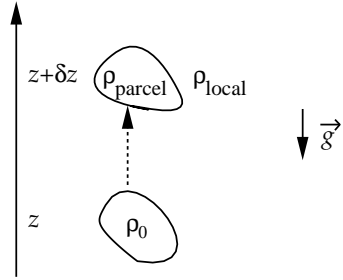}
\end{center}
\caption{As a fluid element is displaced upwards, it will become denser
than its surroundings and will feel a downward restoring force if the 
Brunt-V\"ais\"al\"a frequency is positive, that is if the temperature
gradient is smaller than the local adiabatic gradient.
\label{BVf} }
\end{figure}

\subsubsection{Brunt-V\"ais\"al\"a frequency \label{sec:brunt}}

Let us displace upwards a parcel of fluid from its equilibrium position
in a radiative zone (Fig.~\ref{BVf}). If the displacement over a distance
$\delta z$
is adiabatic and if pressure equilibrium is maintained, 
the density inside the parcel $\rho_{\rm parcel}$ will be larger than the
local density $\rho_{\rm local}$ by an amount 
\beq
\frac{\rho_{\rm parcel} - \rho_{\rm local}}{\rho_0} = \delta z \lp
\left. -\frac {\diff \ln T}{\diff z} \right| _{\rm ad} +
\frac {\diff \ln T}{\diff z} \rp - \delta z 
\frac {\diff \ln \mu}{\diff z} =
\frac{\delta z}{H_P} \lp \nabla_{\rm ad} - \nabla  + \nabla_\mu \rp
\eeq
where $H_P = - \diff z/\diff \ln P$ is the pressure scale height.
We also used the classic notation of stellar modeling $\nabla = \diff \ln T /\diff \ln P$.
The parcel then experiences a restoring force $g \lp \rho_{\rm parcel} - \rho_{\rm local} \rp$,
giving rise to an oscillation characterized by the Brunt-V\"ais\"al\"a frequency
$N$ defined as
\beq
N^2 = N_T^2 + N_\mu^2 = \frac{g}{H_P} \lp \nabla_{\rm ad} - \nabla + \nabla_\mu \rp.
\eeq
The region is locally stable against convection if $N^2$ is positive.

\begin{figure}[t]
\begin{center}
\vspace*{-1cm}
\includegraphics[width=5.5cm]{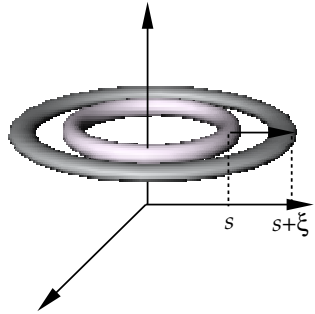}
\vspace*{-0.4cm}
\end{center}
\caption{To take into account the stabilizing effect of rotation,
we expand a torus by an amount $\xi$.
\label{SHinst} }
\end{figure}

\subsubsection{Solberg-H\o iland instability}

The next instability is related to the Coriolis force (Rayleigh~1916,
Taylor~1923). We consider a fluid in {\em cylindrical rotation}. 
In the absence of gravity,
pressure gradients balance centrifugal acceleration
\beq
-\frac{1}{\rho} \frac{\diff P}{\diff s} + s \Omega ^2 =0.
\eeq
Let us displace a fluid torus over some distance
$\xi$ (Fig.~\ref{SHinst}). The net force acting 
on the fluid is proportional to the difference between the 
torus' angular momentum $(s+\xi) \Omega^2_{\rm torus}$ and the equilibrium value 
$(s+\xi) \Omega^2_{s+\xi}$. Since the Coriolis force ensures (specific)
angular momentum $(j=s^2 \Omega)$ conservation, it may be written as
\beq
\lp s+\xi \rp \Omega^2_{\rm torus} - \lp s+\xi \rp \Omega^2_{s+\xi}
= \frac{1}{\lp s+\xi \rp ^3} \lp j_s^2 - j_{s+\xi}^2 \rp \simeq
\frac{1}{s^3} \frac{\diff \left( j^2 \right) }{\diff s} \xi
\eeq
valid to first order in $\xi$. We use this condition to define the
Rayleigh frequency
\beq
N^2_\Omega = \frac{1}{s^3} \frac{\diff}{\diff s} \lp s^2 \Omega \rp ^2
\eeq
which corresponds to a restoring force if positive, that is, if the specific
angular momentum increases outward.

If gravity is taken into account, there are now two restoring forces.
For equatorial radial displacements, the stability condition then reads
\beq
N^2 + N^2_\Omega \ge 0. \label{SHcrit}
\eeq
For general axisymmetric perturbations, the stability criterion is
twofold: \\
\hspace*{0.2cm} $\bullet$ Condition (\ref{SHcrit}) must be satisfied; \\
\hspace*{0.2cm} $\bullet$ Specific angular momentum must increase from pole to equator. \\
The second condition corresponds to applying
the Rayleigh criterion ($N^2_\Omega >0$) to an isentropic surface; 
in the limit of no rotation, the first condition corresponds to the
Schwarzschild criterion. These conditions
have been derived by Solberg (1936) and H\o iland (1941).

\begin{figure}[t]
\begin{center}
\includegraphics[width=6.5cm]{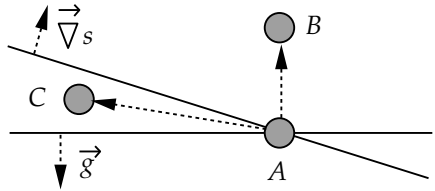}
\end{center}
\caption{In a baroclinic state, equipotentials and isentropic surfaces 
do not coincide. In a displacement from $A$ to $B$, the local entropy
rises, and this state is stable. However, in a displacement from $A$ to $C$,
the local entropy decreases, and this characterizes {\em baroclinic
instabilities}.
\label{baroclinicity} }
\end{figure}

\subsubsection{Baroclinicity}

If the star's rotation state is not cylindrical, it can be shown that
surfaces of constant pressure, density, entropy and gravity do
not coincide\footnote{Baroclinicity could also be caused by uneven
heating (in a binary system )
and this is the case in the Earth's atmosphere.}. 
This is called a {\em baroclinic state} (in opposition to a
{\em barotropic state}). Let us now imagine displacing a fluid element
vertically\footnote{The vertical is defined by the direction of gravity.}
in such a star, going from $A$ to $B$ in Fig.~\ref{baroclinicity}.
Assuming adiabaticity, the fluid will then be denser than its
environment and gravity will restore it to its original position.
However, if the same parcel is moved to $C$, it will now have a lower
density, and gravity will make it rise, leading to an instability,
which is called a {\em baroclinic instability}. However, angular
momentum conservation makes axisymmetric movements towards $C$ impossible
(Knobloch \& Spruit~1982).
In the case of non-axisymmetric disturbances, fluid element exchange
angular momentum (Cowling~1951), allowing instability, which gains energy
by lowering the center of mass of the corresponding fluid.

These non-axisymmetric perturbations have been studied in the
geophysical context using the $\beta$-plane
approximation\footnote{This approximation consist in a local study of the equations
in a tangent plane touching the surface of the sphere at a given latitude. The Coriolis 
force is then expanded linearly around its local value $\beta=2\Omega \cos \theta$.}.
Knobloch \& Spruit (1982) present a summary of these results, and conclude that
instability is possible only for fast rotation that is if
\beq
N^2 < \lp \frac{\diff \Omega}{\diff \ln r} \rp ^2.
\eeq

\subsubsection{GSF instability \label{sec:GSF}}

In stellar interiors, thermal diffusivity $K_T$ is much larger than
viscosity $\nu$ which controls angular momentum diffusion. This weakens
the stabilizing effect of the density stratification (Goldreich
\& Schubert~1967, Fricke~1968). In the case of a vanishingly small
viscosity, Goldreich
\& Schubert (1967) concluded that 
\begin{quote}
... a necessary condition for stability is that the angular momentum per unit
mass be an increasing function of distance from the rotation axis
\end{quote}
which can also be expressed as
\beq
N^2_\Omega \ge 0 ~~~{\rm and}~~~\frac{\partial \Omega}{\partial z}=0,
\eeq
which is similar to the Rayleigh criterion for incompressible, inviscid
fluids (see \eg Chandrasekhar~1961).
This instability which
depends on heat diffusion has a lower growth rate, and this changes the nature
of the instability (it becomes ``diffusive''). Indeed, for thermal diffusivity
to be efficient, the length scale of the perturbation has to be small, and this
reduces the efficiency of the instability.

Fricke (1968) and later Acheson (1978) considered the effect of a finite viscosity and
presented and a modified version of the Solberg-H\o iland criterion for
stability
\beq
\frac{\nu}{K_T} N_T^2 + N^2_\Omega \ge 0
\eeq
together with
%
\beq
\left| s \frac{\partial \Omega}{\partial z} \right| ^2 < 
\frac{\nu}{K_T} N_T^2 ,
\eeq
permitting a small amount of differential rotation in $z$ before
instabilities set in.

\subsubsection{ABCD instability}

If the star is not chemically homogeneous, 
mean molecular weight gradients also contribute to the density 
stratification. Just like the restoring force associated with
the temperature stratification is reduced by the thermal diffusivity,
this contribution is also weakened, but this time by molecular diffusivity 
$K_\mu$. The
Solberg-H\o iland criterion is further modified and becomes
\beq
\frac{\nu}{K_T} N_T^2 + \frac{\nu}{K_\mu} N_\mu^2 + N^2_\Omega \ge 0.
\eeq
Since $K_\mu$ is rather small (of the order of $\nu$), 
this instability likely does not
set in. There is however another possibility. When displacing
a parcel towards $C$, it will come back to its equilibrium position
$A$ due to momentum conservation. However, if thermal diffusion is
large enough, it will then be cooler than its surroundings, and the
parcel will sink further in. This produces growing 
oscillations, a situation called {\em overstability}.
The condition for stability is then (Knobloch \& Spruit~1983)
\beq
\frac{\nu}{K_T} \lp N_T^2  + N_\mu^2 \rp + N^2_\Omega \ge 0.
\eeq

\vspace*{0.2cm}
All criteria described so far are linear, and little work has been
done on the non-linear behavior of the corresponding instabilities, 
implying that we do not know how efficient they are for mixing.

\begin{figure}[t]
\begin{center}
\includegraphics[width=7.5cm]{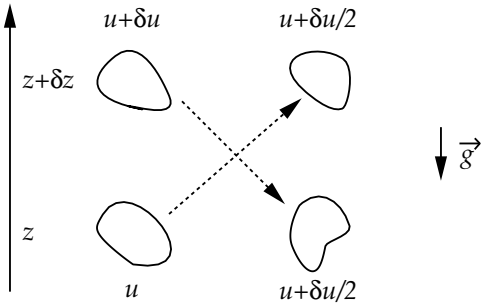}
\end{center}
\caption{To derive the instability condition against shear, 
let us take a blob at level
$z$ with a velocity $u$ and move it upward to the level $z+\delta z$.
Simultaneously, we take a blob at level $z+\delta z$ with a velocity 
$u+\delta u$ and move it down. When doing so, we also homogenize the
velocities that become $u+\delta u/2$ for both blobs. We
gain $1/4 \rho \delta u^2$ in kinetic energy, but lose 
$g \Delta \rho\delta z$ in work against gravity. 
This leads to the Richardson criterion.
\label{shearinst} }
\end{figure}

\subsection{Shear instability \label{sec:shear}}

Rayleigh (1880) derived a linear stability criterion against shear, which
reads simply
\beq
\frac{\diff ^2 u}{\diff r^2}=0
\eeq
everywhere, or in words, there must be no inflection point in
the velocity profile for it to
be stable. In experimental cases however, it is known that turbulence can
develop even in the absence of such an inflection point (that was
shown for example by Wendt (1933) in the study of the flow between
two cylinders). This instability is linked to the presence of
{\em finite amplitude} fluctuations that can appear in the fluid.

To look into the non-linear regime of interest, we can study the
energetics of the problem.
It can be shown that the minimum energy state of a rotating fluid is
solid body rotation. Thus, if the star is rotating differentially, by
homogenizing the velocities, it is possible to extract energy. 
However,
to do so, one must be able to overcome the density stratification.
By comparing both and assuming adiabaticity, one obtains the Richardson 
instability criterion (Fig.~\ref{shearinst})
\beq
Ri \equiv \frac{N^2}{(\diff u/\diff z )^2} < Ri_{\rm crit} = \frac{1}{4}
\eeq
(see \eg Chandrasekhar~1961). This becomes our {\em non-linear} 
condition for instability.

\subsubsection{Dynamical shear instability}

Along isobars, there is no restoring force, and thus, shear is
unstable as soon as horizontal differential rotation is present.
Horizontal shear is thus a dynamical instability, acting on a
dynamical hence quite short time-scale and leading to a large
horizontal turbulent viscosity. A few attempts have been made in order to 
evaluate this value (\eg Richard \& Zahn~1999, Maeder~2003, Mathis, Palacios
\& Zahn~2004), but as of now, there is no reason to prefer one formulation over
the other. It has strong implications on the transport of elements and we
will discuss these in \S~\ref{sec:chem_circu} and \ref{sec:open} (this
is also discussed at length in Talon, Richard \& Michaud~2005).

\subsubsection{Secular shear instability \label{sec:shear}}

In the direction of entropy stratification, both the thermal
and the mean molecular weight stratifications hinder the
growth of the instability. However, thermal diffusion and
horizontal shear act as to reduce the stabilizing effect of the
stratification,
just as in the case of the GSF instability (\S~\ref{sec:GSF}).
We obtain the {\em  instability} criterion 
\beq
\lp \frac{\Gamma}{\Gamma +1} \rp N_T^2 + 
\lp \frac{\Gamma_\mu}{\Gamma_\mu +1} \rp N_\mu^2 < Ri_{\rm crit} 
\lp \frac{\diff u}{\diff z} \rp ^2
\label{eq:shear}
\eeq
where $\Gamma = v\ell/K_T$ and $\Gamma_\mu = v\ell/K_\mu$ (Talon \& Zahn~1997,
see also Maeder~1995). We normally use $Ri_{\rm crit}=1/4$, but this
is open to discussion (see \eg Canuto~1998)
This criterion implies that, the smaller the eddy, the more efficient
thermal diffusivity. In fact, there always exists an eddy that is small
enough so that the instability criterion (\ref{eq:shear}) will be satisfied. 
Turbulent
diffusivity corresponds to the largest eddy satisfying (\ref{eq:shear}); this
is a non-linear description, based on energy considerations. It is
the formulation we shall adopt to describe the vertical shear instability.

\subsection{Meridional circulation \label{sec:MC}}

\begin{figure}[t]
\begin{center}
\includegraphics[width=5.2cm,height=5.2cm]{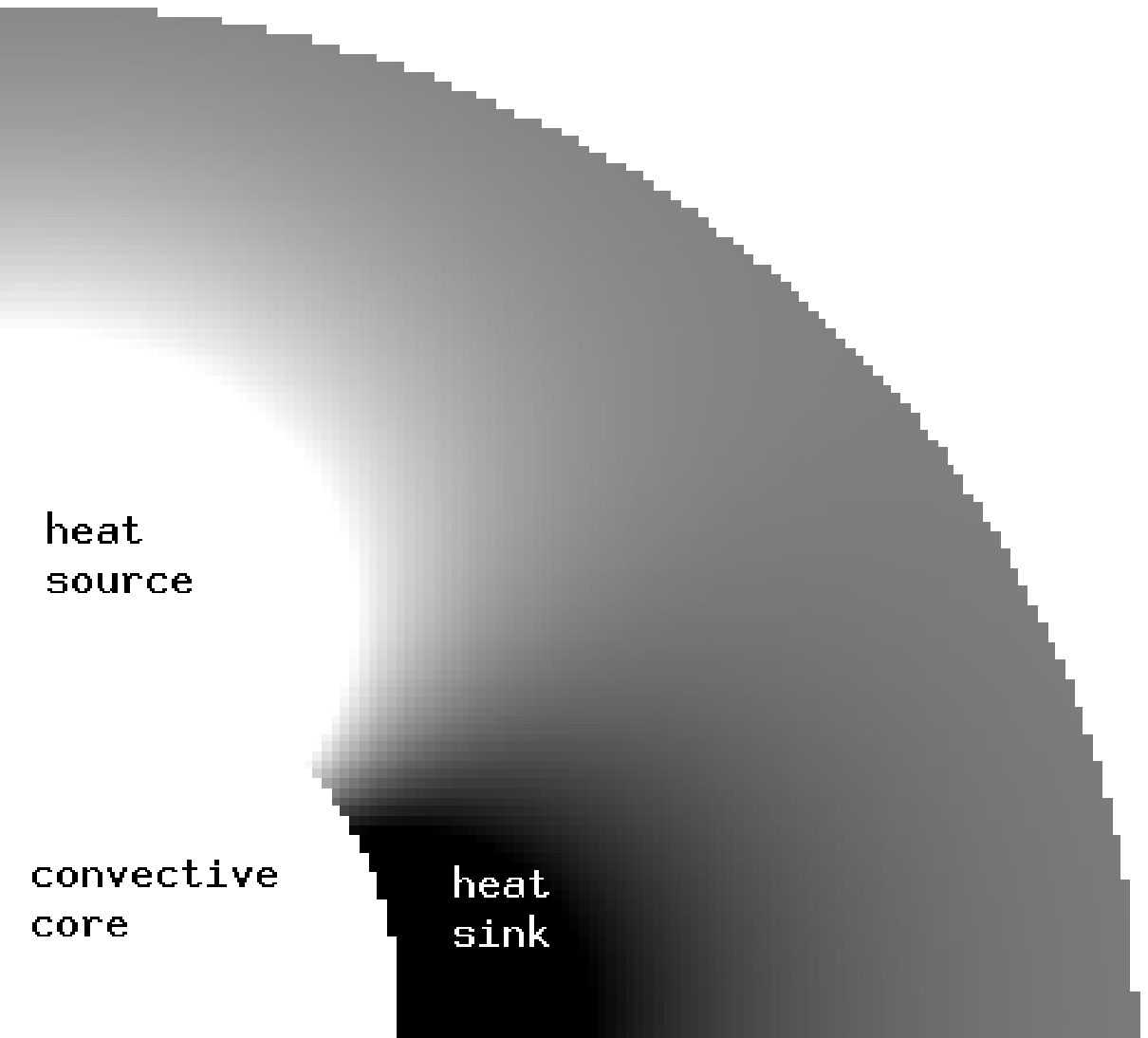}\hspace*{0.8cm}
\includegraphics[width=5.2cm,height=5.2cm]{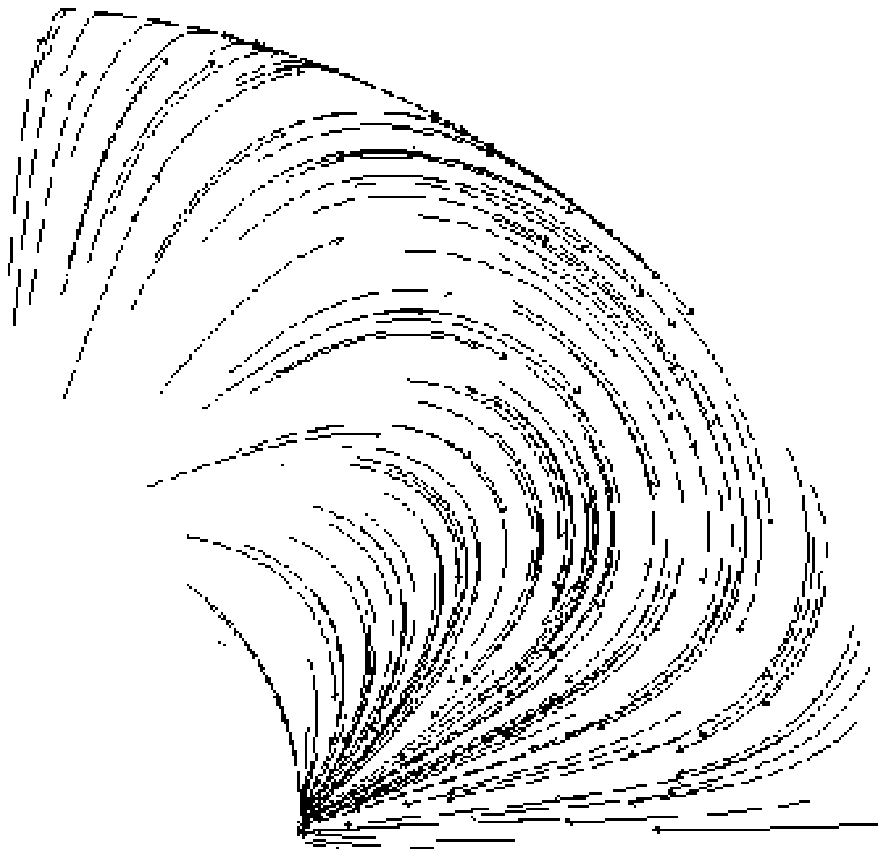}
\vspace*{-0.2cm}
\end{center}
\caption{The traditional approach to meridional circulation consists of
calculating the thermal imbalance for a given rotation state (left, here in
the case of
solid body rotation with an oblateness of 10\%) 
and evaluating the advective flux required to
counter-balance this imbalance (right).
\label{fig:sweet} }
\end{figure}

The Eddington-Sweet meridional circulation is related to the thermal imbalance 
that is in general present in rotating stars. This imbalance was first
pointed out by von~Zeipel~(1924) when he tried to describe thermal equilibrium in a
rotating star in solid body rotation. Since such a star is barotropic, 
equi-potentials and iso-therms coincide, and are ellipsoids. When writing the
divergence of the heat flux
\beq
-\nabla \cdot \vec{F} = \nabla \cdot \lp \chi \nabla T \rp = -\rho \varepsilon,
\eeq
von~Zeipel realized that, while the (horizontal) average of 
$\nabla \cdot \vec{F}$ may
equilibrate the energy generation $\rho \varepsilon$, it is not the case
along an equipotential, 
except if this energy generation has a very special behavior, which
is not realistic (this divergence is shown on the left of Fig.~\ref{fig:sweet}, in
the case $\varepsilon=0$), and this became known as {\em von~Zeipel's paradox}.
Vogt (1925) and Eddington (1925) found a simple solution to the problem, which
consists in considering the advection of entropy $S$ in the heat equation
\beq
\rho T \vec{u} \cdot \vec{\nabla}S = 
\vec{\nabla} \cdot \lp \chi \vec{\nabla} T \rp + \rho \varepsilon ,
\eeq
where $\vec{u}$ is the meridional circulation velocity, which is calculated with
the above equality. This was done initially for the case of solid body rotation
by Sweet (1950) and yields a large scale circulation that rises at the pole and
sinks at the equator.

Sweet's solution is however not fully consistent since this large scale
circulation advects angular momentum as well as entropy. This will modify
the rotation profile, and hence feedback must be treated.
Since this is actually (at least) a 2--D effect, not much progress has been
made in incorporating it into stellar evolution codes until quite recently.

In a first step, models where built by treating the advection of momentum
as a purely diffusive process (Endal \& Sofia~1976, 1978, Pinsonneault et
al.~1989, Langer~1991, these codes are still in use). 

The key ingredient to allow the modeling of meridional circulation 
as a 1--D process
is the assumption of highly anisotropic turbulence,
with the vertical turbulent viscosity $\nu_v$ being much smaller
than the horizontal turbulent viscosity $\nu_h$.
This property
is related to the vertical stratification (Tassoul \& Tassoul~1983,
Zahn~1992) and permits to assume a state of shellular rotation\footnote{Note 
that, in the case of shellular
rotation equi-potentials do not exist, and will be replaced by
isobars.},
{\em i.e.} $\Omega = \Omega(P)$. Conditions for this approximation
to be valid are discussed by Tassoul \& Tassoul (1983), Zahn (1992)
and Charbonneau (1992 - this discussion is based on numerical simulations).

\subsubsection{Heat flux}

Under that assumption, and to first order, it is possible to derive
an equation for the thermal imbalance in a star with a given
(shellular) rotation profile. Then, as in the case of solid body rotation,
meridional circulation is
obtained by stating that advection of entropy $S$ must neutralize this 
thermal imbalance
\beq
\rho T \vec{u} \cdot \vec{\nabla}S = 
\vec{\nabla} \cdot \lp \chi \vec{\nabla} T \rp + \rho \varepsilon - \vec{\nabla}
\cdot \vec{F}_h
\eeq
(this form of the equation also contains a turbulent horizontal heat
flux $\vec{F}_h$,
Maeder \& Zahn~1998). The vertical component of the 
asymptotic circulation velocity is then
\beq
u=\frac{P}{\rho g C_P T}\frac{1}{\lp \nabla _{\rm ad} - \nabla + \nabla _\mu \rp}
\, \lc \frac{L}{M} \lp E_\Omega+ E_\Theta + E_\mu \rp + \frac {TC_P}{\delta} 
\frac{\partial \Theta}{\partial t}\rc \, P_2 \lp \cos \theta \rp
\label{eq:u}
\eeq
(see Maeder \& Zahn~1998 for the complete expression). The 
circulation velocity is related to a term $E_\Omega$ which depends on the
rotation rate, a term $E_\Theta$ which depends on the radial differential
rotation ($\Theta \propto \diff \Omega^2 /\diff r$) 
and a term $E_\mu$ which depends on the horizontal variations
of the mean molecular weight $\mu$. 
In a homogeneous solid body rotating star,
only the first term survives, and one gets Sweet's (1950) classical solution
characterized by a large circulation cell rising along the pole,
with a reversed cell at the surface (Gratton~1945, \"Opik~1951).

Let us note that if convective overshooting is not present, the solution diverges
at the convective boundary where $\nabla _{\rm ad} - \nabla=0$
and viscous boundary layers are then required (Tassoul \& Tassoul~1982).

\begin{figure}[t]
\begin{center}
\includegraphics[width=3.5cm]{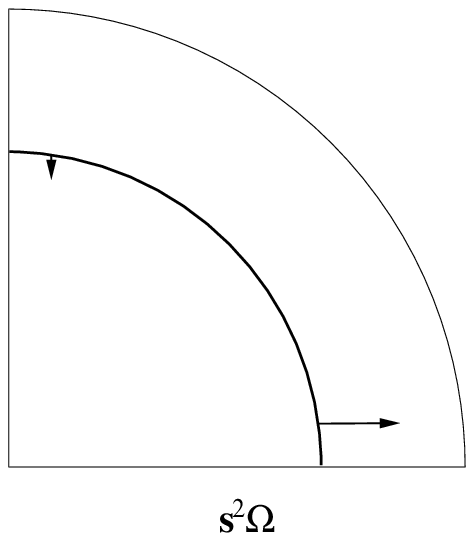}\hspace*{0.8cm}
\includegraphics[width=3.5cm]{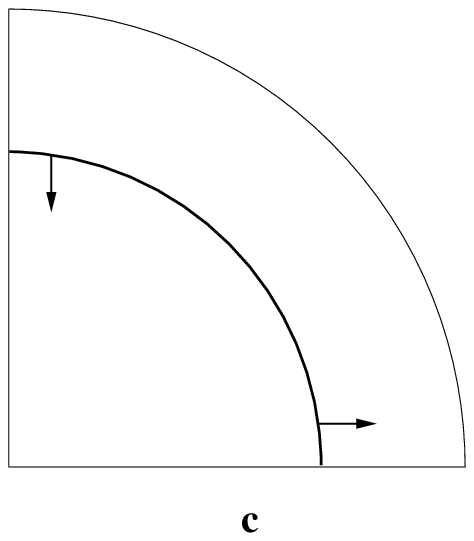}
\vspace*{-0.2cm}
\end{center}
\caption{Meridional circulation has a different effect on
angular momentum and on chemicals. Even when horizontal mixing is strong, the large
scale circulation produces a net flux of $s^2 \Omega$, while there is no
net flux of concentration. In fact, the evolution of $c$ is possible only in the
presence of horizontal inhomogeneities. (From Talon~1997).
\label{compcom} }
\end{figure}

\subsubsection{Angular momentum transport \label{sec:circumom}}

\begin{figure}[t]
\begin{center}
\includegraphics[width=5.8cm]{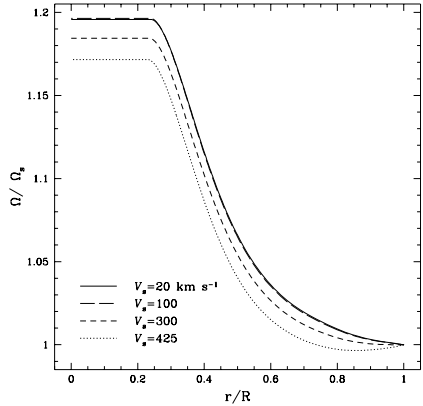}
\vspace*{-0.8cm}
\end{center}
\caption{Equilibrium rotation profile as given by Eq.~(\ref{evoomega})
for various surface velocities in a $9~M_\odot$ main sequence star.
(Figure 1 from Talon et al.~1997.)
\label{fig:eqth} }
\end{figure}

If the only transport processes present are meridional circulation
and hydrodynamical instabilities, the distribution of angular momentum
within the star will evolve according to an advection-diffusion equation
\beq
\rho \frac{\diff}{\diff t} \lc r^2 \Omega \rc = \frac{1}{5r^2} 
\frac{\partial}{\partial r} \lc \rho r^4 \Omega u \rc + \frac{1}{r^4}
\frac{\partial}{\partial r} \lc \rho \nu_v r^4 
\frac{\partial \Omega}{\partial r} \rc 
\label{evoomega}
\eeq
(see Zahn~1992 for details). Note that the first term of this equation
still corresponds to an advective process since, even if $\Omega$ is
homogeneous on isobars, $s^2  \Omega$ is not (see also Fig.~\ref{compcom}).
This will not be the case for chemicals (see \S~\ref{sec:chem_circu}).
Here we use a vertical
turbulent viscosity $\nu_v$ which is attributed to
relevant rotational instabilities detailed in 
\S\,\ref{sec:rot} and \ref{sec:shear}. 
Neglecting the
star's evolution, this equation admits a stationary solution in which
advective transport is counterbalanced by turbulent transport. If no
turbulence is present and if the star is homogeneous, circulation will 
simply stop, resulting in a
rotation profile with a core rotating about 20\% faster 
than the surface
(Urpin et al.~1996, Talon et al.~1997, see also Fig.~\ref{fig:eqth}). 
The equilibrium profile in
$\Theta$ thus depends on the nature and the vigor of rotational
instabilities present in the star. However, when using realistic values
for $\nu_v$, it varies only slightly, and so, meridional circulation is
in general much weaker than predicted by the simple Eddington-Sweet solution.
For any rotation profile, the direction of the circulation can be deduced from
the requirement of reconstructing this equilibrium solution.

If the star is not chemically homogeneous, meridional circulation will
produce horizontal variations of mean molecular weight $\mu$ out of the
vertical chemical stratification. These contribute to the heat flux,
via the equation of state. Equation~(\ref{eq:u}) shows that,
if horizontal variations of mean molecular weight are present, the 
equilibrium profile described in the previous paragraph
exists; however, it now describes the function
$\Theta - \widetilde{\mu}/\mu$ ($\widetilde{\mu}$ refers to
the horizontal fluctuation of $\mu$). 
If rotation evolves solely under the action of meridional circulation
and turbulence, this equilibrium requires an
increase in differential rotation (compared to the homogeneous case)
in order to maintain the asymptotic solution. This effect has been
observed in fully consistent unidimensional calculations
including differential rotation and mean molecular weight gradients
(Talon et al.~1997, Palacios et al.~2003).

If the star is constrained to
rotate as a solid body\footnote{In solid body rotation, no hydrodynamical
instabilities can develop.} and does not evolve, the build up of
horizontal mean molecular weight fluctuations can thus also lead to a 
circulation free state, as originally suggested by Mestel~(1953).

In an evolving stellar model, the required equilibrium profile
also evolves ({\em e.g.} Talon et al.~1997). This leads to
the appearance of
a ``re-adjustment'' circulation, which is added to the asymptotic
circulation. This circulation cannot be avoided, even in
chemically inhomogeneous stars rotating as solid bodies.

\subsubsection{Wind-driven circulation \label{sec:wind}}

If the star's surface is braked via a magnetic torque, as is
the case of Pop~I stars cooler than about 7000~K, the internal distribution
of angular momentum is rapidly moved away form its equilibrium profile,
with large shears developing in the outer regions. This
is the regime of ``wind-driven'' circulation, and the model
presented up to now predicts strong mixing related to such
a state. 

\begin{figure}[t]
\begin{center}
\includegraphics[width=6cm]{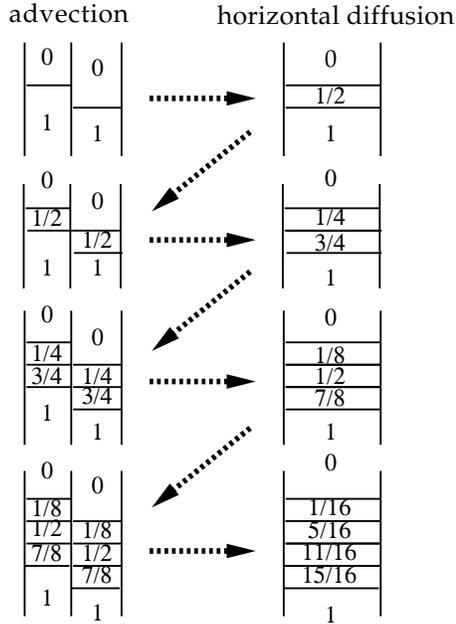}
\vspace*{-0.2cm}
\end{center}
\caption{The combination of advection and horizontal
transport leads to an effective vertical diffusion. In this
cartoon, we are slowly building a discrete version of the error
function, characteristic of a diffusion process.
\label{effdiff} }
\end{figure}

\subsubsection{Chemical transport \label{sec:chem_circu}}

While momentum transport remains a truly advective process when
considering the combined effect of meridional circulation and
horizontal diffusion, it is possible to understand heuristically
that this combination
leads to (vertical) diffusion in the case of chemicals
(Fig.~\ref{effdiff}, see also Chaboyer \& Zahn~1992; this has also been shown
in numerical simulations by Charbonneau~1992).
This is the case because horizontal turbulence continually reduces
horizontal chemical inhomogeneities, impeding efficient advection.
This effect is felt as soon as
\beq
D_h > \left| rU \right| ~~~{\rm and}~~~ D_h/D_v \gta 1000
\eeq
(Charbonneau~1992). In the following, 
we assume that diffusion coefficients are equal to the turbulent
viscosities {\em i.e.} $D_v = \nu_v$ and $D_h = \nu_h$.
The advection + horizontal diffusion + vertical diffusion equation
can then be replaced by a purely diffusive equation, with vertical
turbulent transport and an effective diffusivity depending on
horizontal turbulent transport $D_h$ and advective velocity $u$
\beq
\rho \frac{\diff c}{\diff t} = \rho \dot{c} +
\frac{1}{r^2} \frac{\partial}{\partial r} \lc r^2\rho V_{ip}c \rc +
\frac{1}{r^2} \frac{\partial}{\partial r} 
\lc r^2 \rho \lp D_{\rm eff} + D_v \rp \frac{\partial c}{\partial r} \rc.
\eeq
The effective diffusivity is $D_{\rm eff}=r^2 u^2/\lp 30D_h \rp$ (Chaboyer \&
Zahn~1992); $\dot{c}$ is the nuclear production/destruction rate 
of the species,
and $V_{ip}$ is the atomic diffusion velocity (see \S~\ref{sec:atomic}).

\subsection{Some applications of rotational mixing \label{sec:approt}}

This theory of rotational mixing based on an advective formulation of
meridional circulation and shear instabilities has been applied to a wide
variety of stars. In the case of massive stars, extensive studies have been
conducted by the Geneva group, with great success in explaining problems
such as:\\
\hspace*{0.2cm} $\bullet$ He and N overabundances in O- and early B-type
stars and in super-giants;\\
\hspace*{0.2cm} $\bullet$ the number ratio of red to blue super-giants;\\
\hspace*{0.2cm} $\bullet$ the Wolf-Rayet to O-type stars ratio.\\
Recent results regarding massive stars
are given in Meynet~(2007, this volume).

\begin{figure}[t]
\begin{center}
\includegraphics[width=6cm,angle=-90]{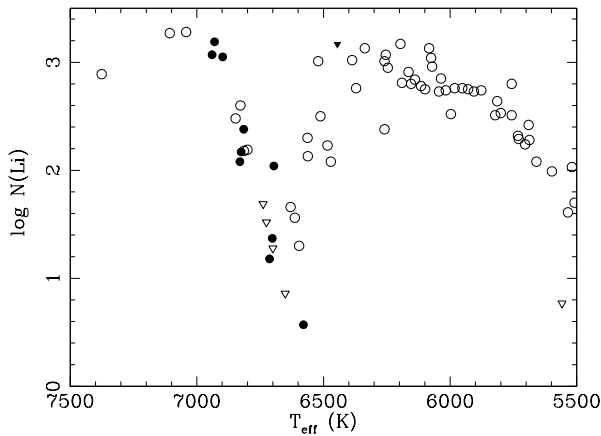}
\end{center}
\caption{Predicted (closed circles) and observed (open circles; inverted open
triangles are upper limits) lithium
abundances in the Hyades around the lithium dip. Theoretical models 
are made with initial velocities of 50, 100 and 150~km\,s$^{-1}$, and final
velocities corresponding to those measured in the Hyades for the corresponding
effective temperature. For cooler stars, lithium depletion remains large ($N({\rm
Li})$ is too small), at
a level incompatible with observations. The inverted filled triangle corresponds
to the prediction of rotational mixing if the inner rotation profile is
artificially assumed to be solid body rotation, with surface braking.
(Adapted from Fig.~6 of Talon \& Charbonnel~1998.)
\label{fig:lidip} }
\end{figure}

In the case of low mass stars, rotational mixing has been applied to the
case of the lithium dip\footnote{The so-called lithium dip, 
discovered by Boesgaard \& Tripicco~(1986) and
observed in all galactic
clusters older than $\approx 300~{\rm Myr}$, refers to a strong lithium depletion in a narrow
region of $\approx 300~{\rm K}$ in effective temperature, 
centered around $T_{\rm eff} \approx 6700~{\rm K}$. Lithium is burned at a
temperature of $\sim 2.5 \times 10^6~{\rm K}$, which, in these stars, is located
well into the radiative zone. As one looks at cooler stars, this burning layer 
is located closer to the base of the convective envelope.}.
In this model, the hot side of the lithium dip corresponds to stars that suffer magnetic
braking (see \S~\ref{sec:wind} and \ref{sec:spindown}); this induces large internal
differential rotation, increasing rotational mixing and thus, surface lithium
depletion. Rotational mixing reproduces well the hot side of the dip (Talon \&
Charbonnel~1998, Palacios et al.~2003, Fig.~\ref{fig:lidip}) but fails for
temperatures below $T_{\rm eff} \lta 6700~{\rm K}$. 

\begin{figure}[t]
\begin{center}
\includegraphics[width=6cm,angle=-90]{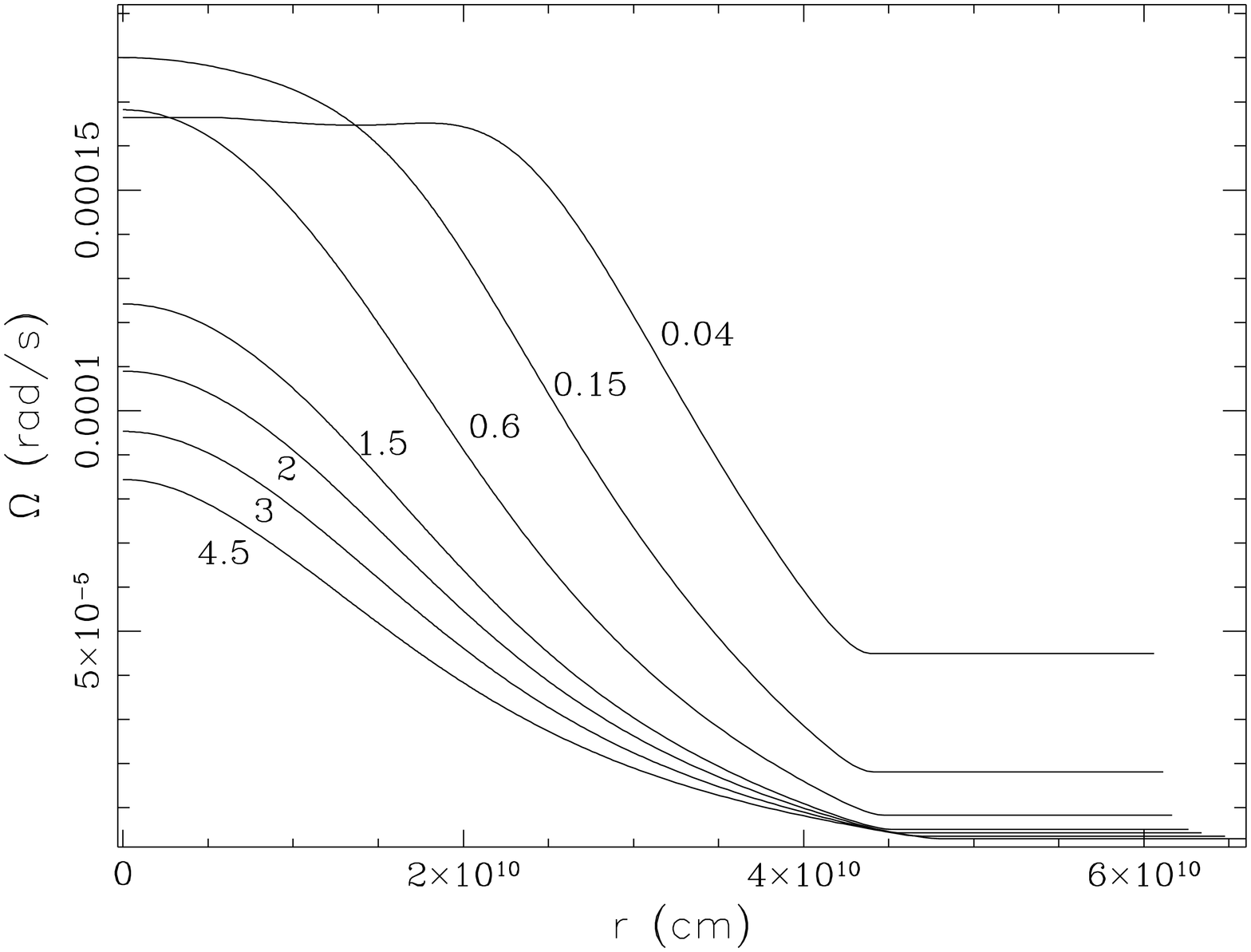}
\end{center}
\caption{Evolution of the solar rotation profile under the effect of rotational
mixing only. The initial rotation velocity is assumed to be 
$100~{\rm km\,s^{-1}}$, and labels
indicate the age in Gyr.
(Figure 11.1 from Talon~1997.)
\label{fig:sun1} }
\end{figure}

Rotational mixing has also been applied to the study of the solar rotation profile.
According to helioseismology, the solar radiative zone is rotating more or less
as a solid body (see \eg Chaplin et al.~1999, Couvidat et al.~2003).
In that case, complete self-consistent calculations predict that radial
differential rotation remains large at the solar age (see Fig.~\ref{fig:sun1}).
This was shown independently by two groups, using 
slightly different approaches to rotational mixing\footnote{The Yale group
uses a diffusive approximation for the meridional circulation instead of the
formulation given in these notes.} and three different stellar
evolution codes (Pinsonneault et al.~1989, Chaboyer et al.~1995, Talon~1997,
Matias \& Zahn~1998). 

These results on low-mass stars indicate that there must be another efficient
transport process for angular momentum active when the surface convection zone
appears; this would explain both the solar rotation profile and the rise of the
lithium abundance on the cool side of the dip (Talon \& Charbonnel~1998).

\subsection{Open problems in rotational mixing \label{sec:open}}

The 1--D modeling of meridional circulation is certainly a very important
step in understanding and testing models of stellar evolution
with rotation. However, one must not forget that it is actually
an advective transport and that unidimensional modeling remains
an approximation. 

The major limitation of the model comes from the invoked horizontal
turbulent diffusion coefficient $D_h$. While in actual models it is
linked to the source of horizontal shear (that is, some function
of $u$) its magnitude remains quite uncertain. Furthermore,
the hypothesis is that its effect will be to smooth out horizontal
fluctuations. This is the case in the Earth's atmosphere
albeit only up to scales that are smaller than the Rossby radius\footnote{This
scale is the one at which Coriolis' force becomes comparable
to vertical stratification.}, which is given by
\beq
L_{\rm Rossby} = \frac{H_P N}{2 \Omega \sin \theta}
\eeq
where $\theta$ is the latitude.
With typical values for $H_P$ and $N$, it is of order 
$L_{\rm Rossby} \simeq (50 R) /(v_{\rm rot} \sin \theta)$, with $v_{\rm rot}$
in ${\rm km\,s^{-1}}$.
This limit will thus become relevant for stars rotating faster than about 
$100~{\rm km\,s^{-1}}$.

\begin{figure}[t]
\begin{center}
\includegraphics[width=5.5cm]{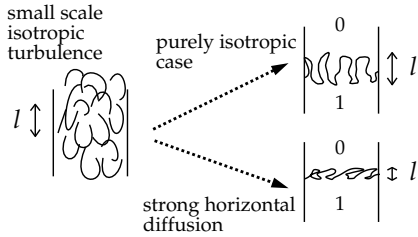}
\vspace*{-2.9cm}
\end{center}
\caption{Strongly anisotropic turbulence reduces the vertical
extent of eddies, leading to a reduction of vertical transport.
\label{anisotropic} }
\end{figure}

Another issue concerns 
the vertical transport of chemicals in the presence of strongly anisotropic
turbulence. As a turbulent eddy is displaced vertically in the
fluid, it looses some of its coherence by being mixed by horizontal
turbulence (see Fig.~\ref{anisotropic}). This effect has been demonstrated in 
numerical simulations (Vincent et al.~1996, Toqu\'e et al.~2006). 
Talon, Richard \& Michaud~(2006) invoked such erosion in the context of AmFm
stars. In \S~\ref{app:dif}, we mentioned that AmFm stars can be understood as
the result of radiative accelerations. However, there is a need for a competing
process to reduce the magnitude of over/underabundances 
predicted in the case of a perfectly stable
envelope. When rotational mixing prescriptions as described here
are applied to such stars,
the associated vertical transport of elements 
is too strong to account for AmFm stars; a reduction of the vertical transport
by horizontal turbulence is required to correctly reproduce the observed
abundances (Talon et al.~2006). However, such a reduction
would degrade the good agreement obtained in the case of massive stars.
Also, the effect of such anisotropic
turbulence on a non-passive
component (\eg angular momentum) remains to be studied. 

A better understanding of the interaction between meridional circulation
and turbulent properties could be obtained via the resolution of
the full Navier-Stokes equations of hydrodynamics in 3--D (see also
 the 2--D, steady, calculations by Garaud~2002). This is
a numerical challenge, as many time-scales are involved in the
problem. 
A first step was taken in that direction by Talon et al.~(2003) who
wrote a code to tackle that problem. Used in direct numerical simulations
with unrealistically large viscosities, it does lead to stationary
solutions. A comparison of these solutions with those of analytical
models should permit a verification of
their validity in the limit of large viscosities.
The goal is then to reduce the viscosities and introduce a sub-scale 
turbulence model
in order to represent unresolved features and to examine how solutions
are modified. 

\section{Magnetic fields}

\subsection{Introduction}

There is a strong connection between rotation and magnetic fields. 
To examine that
relation, let us start by taking a quick look at the equations that govern this interaction.
We first have the induction equation
\beq
\dt{\vec{B}} = \nabla \times \lp \vec{U} \times \vec{B} - \eta \nabla \times \vec{B} \rp,
\label{induction}
\eeq
which describes the action of a velocity field $\vec{U}$ on the magnetic field $\vec{B}$
in the presence of magnetic diffusivity $\eta$. This equation is obtained by
a combination of Amp\`ere's law in the limit of no electric field and of
Faraday's law of induction. We must also consider the momentum 
equation which takes into account the magnetic field by the means of the Lorentz force
\beq
\dt{\vec{U}}+ \lp \vec{U} \cdot \nabla \rp \vec{U} = - \frac{1}{\rho} \nabla P
+ \frac{1}{4\pi \rho} \lp \nabla \times \vec{B} \rp \times \vec{B} + \vec{g}.
\label{B:mom}
\eeq
The relative importance of magnetic field compared to inertial forces
is given by the magnetic Reynolds number ${\rm Re}_m$
\beq
{\rm Re}_m=\frac{ud}{\eta}
\eeq
where $u$ are $d$ are respectively typical velocity and length scales.
The ``stretching'' of field lines is efficient when this number is large.

Here, we will discuss some of the impacts of magnetic fields on stellar structure,
with special focus on their interaction with rotation.

\subsection{Ferraro's law of isorotation}

Probably the best known effect of magnetic fields is to produce solid body
rotation: this is a restatement of Ferraro's theorem (1937), which we will examine
here. To derive this theorem, we first separate our magnetic field into
poloidal and toroidal components
\beq
\vec{B}=\vec{B}_p+ B_\phi  \vec{e}_\phi
\eeq
with $\vec{e}_\phi$ a unit vector in the $\phi$ direction.
We now make various assumptions:\vspace*{0.1cm}\\
\hspace*{0.2cm} $\bullet$ the fluid is perfectly conducting ($\eta=0$);\\
\hspace*{0.2cm} $\bullet$ the fluid is incompressible ($\nabla \cdot \vec{U}=0$);\\
\hspace*{0.2cm} $\bullet$ the solution is axisymmetric;\\
\hspace*{0.2cm} $\bullet$ there is no meridional velocity field;\\
\hspace*{0.2cm} $\bullet$ the solution is stationary.\vspace*{0.1cm} \\
The induction equation for the $B_\phi$ component then reads
\beqan
\dt{B_\phi} &=& \lc \nabla \times \lp \vec{U} \times \vec{B} \rp 
\rc \cdot {\vec{e}_\phi} \nonumber \\
&=& \lc \vec{U} \lp \nabla \cdot \vec{B} \rp -\vec{B} \lp \nabla \cdot \vec{U} \rp  
+ \lp \vec{B} \cdot \nabla \rp \vec{U} - \lp \vec{U} \cdot \nabla \rp \vec{B}
\rc \cdot {\vec{e}_\phi} \nonumber \\
&=& \lc \lp \vec{B}_p \cdot \nabla \rp U_\phi{\vec{e}_\phi}
- \lp U_\phi{\vec{e}_\phi} \cdot \nabla \rp \vec{B}\rc \cdot {\vec{e}_\phi}
 \label{inducphi}
\eeqan
This equation shows how differential rotation 
produces a toroidal field from a purely poloidal field.
In order to obtain a stationary state, we require
\beq
\lp \vec{B}_p \cdot \nabla \rp \Omega=0 \label{ferraro}
\eeq
with $U_\phi=r \sin \theta \,\Omega$.
This equation indicates that, under equilibrium conditions, fluid parcels that
are located along the same poloidal field line must have the same angular velocity.

\subsection{Alfv\'en speed}

It is not sufficient to know what the equilibrium state is, one must also evaluate the
timescale required to reach this equilibrium. Here, we want to know how a field line,
that is stretched perpendicularly, responds. This gives rise to an Alfv\'en wave,
and the speed of this propagation is thus the Alfv\'en speed $v_A$\footnote{This
velocity is obtained by coupling linearized versions of Eqs.~(\ref{induction})
and (\ref{B:mom}).}
\beq
v_A=B_0 \lp 4 \pi \rho \rp^{-1/2}
\eeq
(Alfv\'en~1942).
In the star, shear will propagate over a stellar radius
$R$ in an Alfv\'en time
\beq
\tau_A = \frac{R}{v_A}
\eeq
which also defines the Alfv\'en frequency
\beq
\omega_A=\frac{1}{\tau_A}=\frac{v_A}{R}.
\eeq
Let us mention that, in the absence of dissipation, this shear will simply bounce back
and forth, and iso-rotation along poloidal field lines would not be obtained
(see e.g. Mestel \& Weiss~1987).
Furthermore, in order to reach solid body rotation, diffusion is also required to
``connect'' the fluid velocity between various field lines. These two effects
contribute to a significant increase of the timescale 
required to actually reach solid body rotation.

\subsection{Solar spin-down \label{sec:spindown}}

We already mentioned in \S~\ref{sec:wind} the problem of magnetic braking
that is observed in all stars that have a ``significant'' surface convection zone,
in practice all stars cooler than $T_{\rm eff} \lta 7000~{\rm K}$.
This can be understood in terms of mass loss in a magnetized wind, as originally
suggested by Schatzman~(1962) and Weber \& Davis~(1967): it is related to the fact
that, when magnetic energy is larger than the wind's kinetic energy, matter is
bound to co-rotate with the star even at large distances, leading to a large angular
momentum loss even for a modest mass loss. An order of magnitude calculation is as
follows.
The surface solar dipolar field is
\beq
B_{0r} \approx 1~{\rm G}.
\eeq
Magnetic flux conservation imposes
\beq
r^2 B_r = R_\odot^2 B_{0r}
\eeq
while mass conservation leads to
\beq
\rho u_r r^2 = \rho_0 u_{0r} R_\odot^2
\eeq
where $u_r$ is the radial velocity in the wind for the mass loss.
The comparison of the kinetic energy to the magnetic energy is given by the
Alfv\'enic Mach number 
\beq
M_A^2 = \frac{4\pi \rho u_r^2}{B_r^2} = \frac{1}{\rho}\frac{4\pi \rho_0^2
u_{0r}^2}{B_{0r}^2} \equiv \frac{\rho_A}{\rho}
\eeq
which we use to define the Alfv\'en density $\rho_A$, corresponding to the
density at which $M_A=1$. If the mass loss velocity $u_{0r}$ is known, one can evaluate
this density. Using the values at the Earth orbit, $u_E \simeq 400~{\rm
km\,s^{-1}}$ and $\rho_E \simeq 12 \times 10^{-24}~{\rm g\,cm^{-3}}$ leads
to a density of
\beq
\rho_A \simeq 5 \times 10^{-21} ~{\rm g\,cm^{-3}}.
\eeq
Building a steady wind model similar to Parker's (1960), Weber \& Davis (1967) concluded
that the corresponding Alfv\'en radius is
\beq
r_A \simeq 24 R_\odot.
\eeq
The wind must co-rotate with the magnetic field up to that distance, explaining the
efficiency of the surface braking. A more complete description of this process which
also discusses the role of various field configurations is given by Kawaler~(1988, see
also references therein). 
This produces surface braking of the star. Through convective motions, it is expected
that this braking affects the whole convection zone. The problem of the transport of
angular momentum in the radiative zone remains.

\subsection{The solar rotation}

Another interesting application of Ferraro's law lies in the spin down of the solar
radiative zone. As already mentioned (\S~\ref{sec:approt}), helioseismology concludes
that the the Sun's deep interior is rotating more or less rigidly, and rotational mixing
alone is not able to provide an explanation of this feature. An obvious solution
to this problem is to invoke a fossil magnetic field, as described at length
by Mestel \& Weiss~(1987). Charbonneau \& MacGregor (1993, see also Barnes, Charbonneau,
\& MacGregor~1999) 
studied this effect in a 2--D solar model, assuming a {\em fixed} poloidal
configuration. They applied surface braking to a star originally rotating at
$50~\Omega_\odot$. 
In these simulations, as the surface is spun-down, the poloidal field is
sheared; this builds up a toroidal field that will then, through the Lorentz
force, oppose the shear. This toroidal field grows until the torque exerted by
magnetic stresses 
\beq
\tau = r^3 \left< B_r B_\phi \right>
\eeq
is strong enough to oppose the wind-mediated torque. At this stage, the core and
the envelope fully couple, and spin-down together, with the overall
differential rotation remaining small. This occurs for all values of the
poloidal field, although the time-scale for this re-coupling to occur depends
on $B_r$.
Several configurations have been tested for the poloidal field, with the field
lines anchored in the convection zone (configurations
D1 and D2 of their models), or not (configuration D4) or barely touching 
(configuration D3).

Initial calculations were performed with the convection zone rotating as a solid
body. When latitudinal differential rotation in the convection zone 
is present, due to Ferraro's law, it imprints itself in the interior 
(MacGregor \& Charbonneau~1999). Agreement with helioseismic inversion require
that the field be totally contained in the radiative zone (configuration D4).

\begin{figure}[t]
\begin{center}
\includegraphics[width=9.5cm]{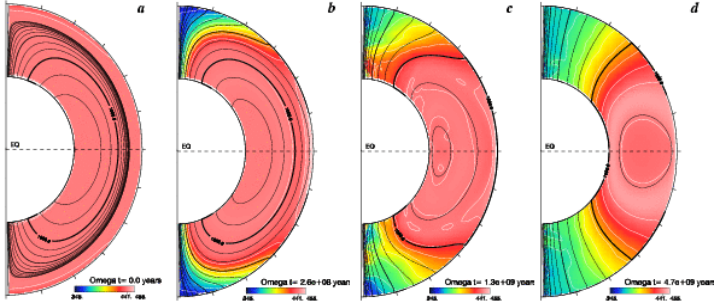}
\end{center}
\caption{Temporal evolution, in a 3--D MHD model of the solar radiative zone, 
of the
angular velocity $\Omega$ (color contours and white lines) and the mean
axisymmetric poloidal field (superimposed black lines), which initially is buried
deeply in the interior (Fig.~4 of Brun \& Zahn 2006). 
{\bf a-d)} sequence spanning 4.7 Gyr
(in solar equivalent units) of the dipolar magnetic field in the presence of
rotation and shear. The field lines gradually connect with the imposed top shear,
thereby enforcing differential rotation at all depths above latitudes greater
than $\approx 40^\circ$, at the (equivalent) age of the Sun. Note also in frames
{\bf c-d} the distortion  near the poles of the lines of constant $\Omega$, which
is associated with the Tayler instability (see \S~\ref{sec:TS}). 
\label{B_brun} }
\end{figure}

Recent 3--D MHD numerical simulations in which the fossil poloidal field is allowed to
evolve have been performed to study this mechanism in a fully consistent way,
using the ASH code (Brun \& Zahn~2006 and references therein).
In all cases studied by these authors, the fossil field is found to diffuse
outwards and inexorably connects with the outer convection zone, imprinting its
differential rotation to the radiative interior. Contrary to the expectation of
Gough \& McIntyre~(1998), the meridional flow that develops between the base of
the convection zone and the dipole poloidal field is not able to stop its
progression. It remains to be seem whether things would change if lower (more
realistic) values of the magnetic diffusivity, thermal diffusivity and
viscosity were used (this is currently out of reach numerically).

\subsection{Magnetic stability}

Another issue that needs concern stellar astrophysicists regarding magnetic 
fields
is that of the magnetic stability. This is well illustrated by the following
quote from Braithwaite \& Spruit (2004):
\begin{quote}
``Although there have been educated guesses as to what shape a field in a stable
equilibrium might have, all configurations studied with the analytic methods
available so far were found to be unstable on the Alfv\'en timescale; it has
been impossible to prove stability for even a single field configuration.''
\end{quote}
However, while the presence of a large scale magnetic field in the solar
radiative zone remains unproven, such fields are actually measured in several
stars, as \eg in Ap stars. Braithwaite \& Spruit made numerical
simulations of such stars, in order to understand what the structure of their
field should look like. In their 3--D simulation, they model a non-rotating $n=3$
polytrope, in which they introduced an initial random magnetic field. The total
energy $E_B$ in that initial field is assumed rather small, that is
$E_B \approx 0.01 E_T$, with $E_T$ the thermal energy of the star.

They found that, after a few Alfv\'en times, 
it was indeed possible to find a stable magnetic structure,
which has the following properties: \vspace*{0.1cm}\\
\hspace*{0.2cm} $\bullet$ the field is approximately axisymmetric;\\
\hspace*{0.2cm} $\bullet$ the poloidal field has the same amplitude as the
toroidal field;\\
\hspace*{0.2cm} $\bullet$ magnetic helicity is conserved.
\vspace*{0.1cm} \\
The properties of their magnetic field are in close agreement with those
actually measured in Ap stars, giving strength to the idea that it is of
fossil rather than dynamo origin.

\subsection{The Tayler-Spruit instability \label{sec:TS}}

The issue of magnetic stability in rotating stars
has been studied theoretically 
in great details by several authors, and an extensive
review of these can be found in Tayler~(1973) and Pitts \& Tayler~(1985).
These linear stability study are performed using perturbations of the equations
for the conservation of mass, momentum (Eq.~\ref{B:mom})
and energy and the magnetic induction
equation (Eq.~\ref{induction}) and verifying whether a lower energy state
exists. Actual calculations are time consuming, and have
to be made for each magnetic configuration.
The relative efficiency of these instabilities was reexamined by Spruit~(1999),
and he concluded that, the {\em pinch-type instability}, which is now
more widely known as the {\em Tayler-Spruit instability} in the stellar
community, was the
first to set in. We will review this process here.

\begin{figure}[t]
\begin{center}
\includegraphics[width=5cm]{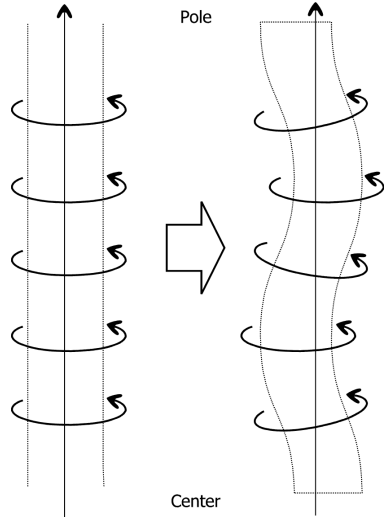}
\end{center}
\caption{In the kink instability, toroidal field loops around
the rotation axis move (mostly horizontally) relative to each other. Adapted from Tayler~(1973).
\label{kink} }
\end{figure}

Starting from a seed poloidal field $B_r$, differential rotation winds it up 
and generates a mainly azimuthal field $B_\phi \gg B_r$. The most unstable mode
for this horizontal field is then a non-axisymmetric 
``kink'' mode, in which the displacement is mainly horizontal (to
minimize work against the density stratification) and incompressible (to minimize
work against pressure) (see Fig.~\ref{kink}). Together, these two conditions
imply 
\beq
l_h/l_r \approx \xi_h/\xi_r \gg 1
\eeq
where $l_h$ and $l_r$ are respectively the horizontal and vertical length scales
and $\xi_h$ and $\xi_r$ are the horizontal and vertical displacements.
Ignoring thermal diffusivity, the work against buoyancy is
\beq
W=\frac{1}{2} \rho \xi_r^2 N^2
\eeq
while the kinetic energy of the displacement is
\beq
K= \frac{1}{2} \rho \xi^2 \omega_A^2 \simeq 
\frac{1}{2} \rho \xi_h^2 \omega_A^2 
\eeq
which yields the instability condition
\beq
l_r < \frac{r \omega_A}{N} \label{l_r}
\eeq
assuming $l_h \simeq r$. This limits the vertical amplitude of the instability.
However, magnetic diffusivity damps those small scales. The characteristic
time-scale for magnetic diffusion $t_d=l_r^2/\eta$ has to be larger than
the growth time of the instability which, in the limit of strong Coriolis force
is of order 
\beq
\sigma_B = \omega_A^2/\Omega, \label{growthBphi}
\eeq 
for the instability to
experience a net growth. One finds another condition for
the vertical scale
\beq
l_r^2 > \frac{\eta \Omega}{\omega_A^2}. \label{mag_diff}
\eeq
The combination of these two conditions yields a criterion for magnetic
instability
\beq
\frac{\omega_A}{\Omega} > \lp \frac{N}{\Omega} \rp ^{1/2} \lp \frac{\eta}
{r^2 \Omega} \rp ^{1/4}. \label{cond1no}
\eeq
This condition, valid in the limit of zero thermal diffusivity $K_T$, was
originally derived by Acheson~(1978). It would correspond to
a region where $N$ is dominated by mean molecular weight gradients.
When $K_T$ is taken into account (or in regions where $N$ is dominated by the
thermal stratification),
the required work against gravity is reduced by $K_T/\eta$, as in the
case of hydrodynamic instabilities (\S~\ref{sec:GSF} and \ref{sec:shear}).
The modified instability criterion then reads
\beq
\frac{\omega_A}{\Omega} > \lp \frac{N}{\Omega} \rp ^{1/2} \lp \frac{K_T}
{r^2 \Omega} \rp ^{1/4} \lp \frac{\eta}{K_T} \rp ^{1/2} \label{cond2no}
\eeq
(Spruit~1999). For a sufficiently large magnetic field, there is a range
of radial scales that are unstable with the largest unstable scale limited
by stratification (Eq.~\ref{l_r}) and the smallest scale, by
magnetic diffusivity (Eq.~\ref{mag_diff}). These small scales produce
{\em turbulent} magnetic diffusivity given by the equality in 
Eqs.~(\ref{cond1no}) and (\ref{cond2no})
\beq
\eta_{\rm e0} = r^2 \Omega \lp \frac{\omega_A}{\Omega} \rp ^4 \lp
\frac{\Omega}{N} \rp ^2 ~~~{\rm or}~~~
\eta_{\rm e1} = r^2 \Omega \lp \frac{\omega_A}{\Omega} \rp ^2 \lp
\frac{\Omega}{N} \rp ^{1/2}  \lp \frac{K_T}{r^2N} \rp ^{1/2}
\eeq
in cases with (0) and without (1) thermal diffusivity.

Spruit~(2002) made further assumptions to produce a model that can be 
used in stellar evolution codes. Equating the growth-rate of $B_\phi$ (Eq.~\ref{growthBphi})
with the amplification time of the Tayler-Spruit instability
\beq
\tau=\frac{N}{\omega_A \Omega q}~~~{\rm with}~
q=\left| \frac{\partial \ln \Omega}{\partial \ln r} \right| ,
\eeq 
he obtains a relation between the Alfv\'en frequency (or equivalently, the amplitude of
the magnetic field) and the differential rotation
\beq
\frac{\omega_A}{\Omega} = q \frac{\Omega}{N}
\eeq
valid in the absence of thermal diffusivity and
\beq
\frac{\omega_A}{\Omega} = q^{1/2} \lp \frac{\Omega}{N} \rp^{1/8} 
\lp \frac{K_T}{r^2 N} \rp ^{1/8}
\eeq
when $K_T$ cannot be neglected.

Feed back to the rotation profile occurs through magnetic stresses. Still assuming $B_\phi \gg B_r$, one gets
\beq
B_\phi = \lp 4 \pi \rho\rp^{1/2} r \omega_A ~~~{\rm and}~~~B_r = \frac{l_r}{l_h}B_\phi
\approx \frac{l_r}{r}B_\phi
\eeq
($l_r$ is obtained by assuming the equality in condition~\ref{l_r}). Magnetic stresses are then of the form
\beq
S_0 = \frac{1}{4\pi} B_r B_\phi 
= \rho r^2 \Omega^2 q^3 \lp \frac{\Omega}{N} \rp^ 4 ~~~{\rm or}~~~
S_1 = \rho r^2 \Omega^2 q \lp \frac{\Omega}{N} \rp^ {1/2} \lp \frac{K_T}
{r^2 N} \rp ^{1/2}
\eeq
and this corresponds to a viscosity for the vertical transport of angular momentum
\beq
\nu_{\rm TS0} = \frac{S}{\rho q \Omega} = r^2 \Omega q^2 \lp \frac{\Omega}{N} \rp^ 4 ~~~{\rm or}~~~
\nu_{\rm TS1} = r^2 \Omega \lp \frac{\Omega}{N} \rp^ {1/2} \lp \frac{K_T}
{r^2 N} \rp ^{1/2}
\eeq
with indexes (0) and (1) representing cases with and without thermal diffusivity
(Spruit~2002). Maeder \& Meynet~(2004) made a more general formulation, which takes into
account intermediate regimes.

\begin{figure}[t]
\begin{center}
\includegraphics[width=9.5cm]{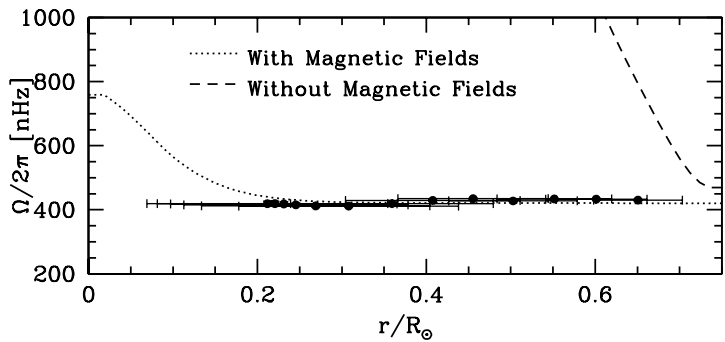}
\vspace*{-0.5cm}
\end{center}
\caption{Rotation profile at the solar age for a model with meridional circulation and
shear instability (dashed line) and with also the contribution of the Tayler-Spruit 
instability (dotted line). The initial velocity is $50~{\rm km\,s^{-1}}$, and surface braking
follows Kawaler~(1988). Dots with error bars correspond to the helioseismic solar rotation
(Couvidat et al.~2003). From Eggenberger et al.~(2005).
\label{eggenberger} }
\end{figure}

\subsubsection{Applications to stars}

This model has been applied by the Geneva group to a variety of stellar models. This process is highly efficient in reducing differential rotation. In the case of massive stars, rotation gradients are strongly reduced by the Tayler-Spruit instability 
(Maeder \& Meynet~2004). Shear instabilities become much less important in this case, while meridional circulation is revived, since the rotation profile remains far from the one required by thermal equilibrium (\S~\ref{sec:circumom}).
These results are discussed at length by Meynet~(2007, this volume).

In the case of the Sun, the Tayler-Spruit instability also strongly reduces differential
rotation. This was tested in an evolutionary model of a $1~M_\odot$ model, in which
are also included meridional circulation and shear instability (Eggenberger, Maeder
\& Meynet~2005). This study confirms the capability of this process to reproduce the Sun's flat rotation profile (see Fig.~\ref{eggenberger}).

\subsection{Open problems in the coupling of magnetic fields and rotation}

The existence of the Tayler-Spruit instability has been observed in numerical 
simulations
(Braithwaite~2006). However, the interaction of this instability with the 
star's
rotation with the mechanism described above implicitly assumes 
the regeneration of
a large scale, axisymmetric poloidal field from which the toroidal field 
can be generated. 
However, the ``kink'' instability described here corresponds to a global 
$m=1$ mode, and so
generates a poloidal field with a similar angular dependence, and 
not the axisymmetric
field that is required in the model. This was also 
shown in numerical simulations
by Zahn, Mathis \& Brun~(2007). In their simulations, the dynamo for 
the poloidal field assumed by
Spruit~(2002) is not present. It remains to be seen if it 
would occur in conditions
with smaller values of $\eta$ and $\nu$ as expected in stellar interiors.

Another approach to the interaction of magnetic fields and rotation consists in adding
the Lorentz force in the equations for meridional circulation (Mathis \& Zahn 2005).
This leads to a complex set of 16 coupled differential equations that has not yet been included
in actual stellar evolution calculations.

\section{Internal gravity waves}

\subsection{Introduction}
A stratified fluid is one in which the density changes 
with height. In the oceans, 
such stratifications are stable if 
there is a continuous decrease in density with height; that is, lighter fluid is
always located above heavier fluid. In the Earth's atmosphere and in stars, 
one has to remove the adiabatic temperature gradient when making this comparison. This gives rise to the well known Schwarzschild criterion.

When a stable stratification is disturbed by a perturbation, 
buoyancy and gravity tend to restore the fluid to its previous equilibrium.
For example, when a fluid parcel is displaced upward
into a less dense region, gravity pulls it back down, below its equilibrium position
and thus, into a denser region.
Buoyancy takes over and pushes it upward. The fluid parcel ``overshoots'' to
the less dense region and the cycle repeats. 
This oscillatory behavior forms what is called an
internal gravity wave.

In the Earth's atmosphere, wave induced momentum transport is a key element in 
the understanding of several phenomena; the most famous is the quasi-biennial 
oscillation (or QBO), which 
is an alternating pattern of westerly and easterly mean zonal winds observed 
in the stratosphere close to the equator. 

In astrophysics,
internal gravity waves have initially been invoked as a source of mixing
for elements induced by a variety of mechanisms
(Press~1981, Garc\'{\i}a L\'opez \& Spruit~1991,
Schatzman~1993, Montalb\'an~1994,
Montalb\'an \& Schatzman~1996,~2000, Young et al.~2003, see \S\,\ref{sec:part}).
Ando~(1986) studied the transport of momentum
associated with standing gravity waves; he showed how angular momentum 
redistribution by these waves may increase
the surface velocity to induce episodic mass-loss in Be stars
(see also Lee~2006).
He was the first to clearly state (in the stellar context)
the fact that waves carry angular momentum
from the region where they are excited to the region where they are dissipated.
Goldreich \& Nicholson~(1989, see also Zahn~1975) later invoked gravity waves in order to explain
the evolution of the velocity of binary stars, producing  
synchronization that proceeds
from the surface to the core. Traveling internal gravity 
waves have since been invoked as an
important process in the redistribution of angular momentum
in single stars spun down by a magnetic torque (Schatzman~1993, 
Kumar \& Quataert~1997, Zahn, Talon, \& Matias~1997, 
Talon, Kumar, \& Zahn~2002).


\subsection{Properties of internal gravity waves \label{sec:prop}}

Internal gravity waves (IGWs) propagate
due to buoyancy forces. Contrary to surface waves which are confined 
to a fluid's interface, internal gravity waves are able to propagate both
horizontally and vertically through the fluid. They exist in stratified environments
and transport energy and momentum.

\begin{figure}
\begin{center}
  \includegraphics[width=7cm]{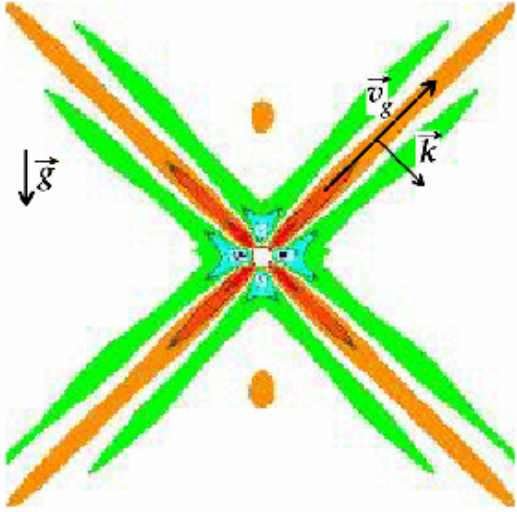}
  \vspace*{-0.5cm}
\end{center}
  \caption{Numerical simulation of an internal gravity wave excited by an oscillatory 
  forced patch. Colors show density fluctuations. {\em Credit: Bruce R. Sutherland, U. of Alberta.} \label{fig:groupvel}}
\end{figure}

Because their restoring force is buoyancy, these waves are anisotropic
with respect to 
``vertical'' and ``horizontal'' orientations. This implies that, contrary to the case of a pressure wave, a point source does not produce an isotropic wave.
Their frequency $\sigma$ has to be lower than the 
Brunt-V\"ais\"al\"a
frequency $N$, which is the natural oscillation frequency of a displaced element in a stratified region. We thus have
\beq
0 < \sigma < N
\eeq
($N$ is defined in \S~\ref{sec:brunt}).
In stars, $N_T$ is typically of order $10^{-3}\,{\rm s}^{-1}$.
A surprising property of internal waves is that their group velocity $\vec{v}_g$,
which corresponds to the transport of energy is perpendicular to the phase 
velocity $\vec{v}_p$ and to the wave number\footnote{This can be demonstrated
using the real part of the dispersion relation (\ref{reldisp}).} $\vec{k}$. 
This is illustrated in 
Fig.~\ref{fig:groupvel}.

Let us now take a more formal look at the equations which
describe those waves in a differentially rotating star\footnote{Our 
development follows Press~(1981), Goldreich \& Nicholson~(1989)
and Zahn et al.~(1997). See also Unno et al.~(1989) for details
regarding the derivation of the wave equation.}. We use
spherical coordinates ($r$, $\theta$,  $\phi$). 
$\vec e_z$ is the unit vector along the rotation axis, and $\vec e_\phi$ that in the
azimuthal direction. We assume that the angular velocity $\Omega$
depends only on depth (just as in \S~\ref{sec:MC}).
In an inertial frame, the total velocity $\vec U$ is
\beq
\vec U (r,\theta, \phi, t)  = \Omega (r) \, \vec e_z  \times \vec r +
\vec u (r,\theta, \phi, t), 
\label{totv}
\eeq
where $\vec u$ is the velocity associated with the wave.
The equation of motion reads
\beq
{\diff\vec u  \over \diff t}
+ \Bigl\{ 2 \Omega \,\vec e_z \times \vec u + \vec e_\phi\,
r \sin \theta \lp \vec u
\cdot \vec \nabla \Omega \rp \Bigr\} 
= - {1 \over \rho} \vec \nabla P' + {\rho ' \over \rho} \vec g  ,
\label{motion}
\eeq
with
\beq
{\diff \over \diff t} = \left({\partial \over \partial t} +
\Omega {\partial \over \partial \phi} \right) ,
\label{doppler}
\eeq
and where $P'$ and $\rho'$ are the Eulerian
perturbations of pressure and density respectively.
We used the Cowling~(1941) approximation and neglected fluctuations 
of the gravitational potential. Viscosity is also 
ignored. We also consider the continuity equation
\beq
{\diff \rho' \over \diff t}
 + \vec \nabla \cdot \rho \vec u = 0,
\label{cont}
\eeq
and the thermodynamic relation between $\rho'$ and $P'$ which in the adiabatic 
limit reads
\beq
{\rho' \over \rho} - {1 \over \Gamma_1}{P' \over P}
- \frac{N^2}{g} \xi_r = 0 ,
\label{adiab}
\eeq
$\Gamma_1$ being the adiabatic exponent and $\xi_r$ the radial part of
the displacement and which is related to the wave velocity by $u_r = i \sigma \xi_r$.

We will proceed with a further simplification. Neglecting 
the terms in curly brackets in the equation of motion
(\ref{motion}), we treat waves as if they were pure gravity waves which
are not modified by the Coriolis acceleration, but just feel the
entrainment by the differential rotation\footnote{The case of the gravito-inertial wave
can be treated using the so-called {\em traditional approximation}
which consists of neglecting the radial component of the Coriolis acceleration in the
equation of motion. In that case, the horizontal structure can be described by 
Hough functions. This is discussed by Lee \& Saio~(1997) and Townsend~(2003). 
The dispersion relation for
gravito-inertial-Alfv\'en waves is discussed in Kumar, Talon, \& Zahn~(1999).}.
In this case, we recover the equations that describe
internal waves in a non-rotating star up to replacing the derivative
with respect to time by (\ref{doppler}), see \eg Unno et al.~(1989).
Solutions are then separable with respect to $r$, $\theta$, $\phi$ and $t$ and
are of the form
\beq
u_r\lp r, \theta, \phi, t\rp = u_v \lp r\rp\, P_\ell^m\lp\cos \theta\rp\,
\exp i \lp \sigma t - m \phi \rp,
%
%
\label{vvert}
\eeq
where
\beq
\sigma(r) = \sigma_0 - m \Omega(r) ,
\eeq
with $\sigma_0$ being the frequency in the inertial frame and $P_\ell^m$ is the 
associated Legendre polynomial of order $\ell$ and azimuthal number $m$.
In the internal waves low frequency range which is considered here,
terms of order $\sigma ^2$ can be neglected; this corresponds to the anelastic
approximation. 
The function $\Psi (r) = \rho^{1\over2} r^2 u_v$ then obeys the second order
equation 
\beq
\frac{\diff^2 \Psi}{\diff r^2} + \lp \frac{N^2}{\sigma^2} -1 \rp
{\ell(\ell+1)\over r^2} \Psi = 0 
\label{secorder}
\eeq
and the dispersion relation is simply
\beq
k^2 \sigma ^2 + N^2 k_h^2 =0
\eeq
(Unno et al.~1989).
Here, we neglected a right hand side term of order $H_\rho ^{-2}$
which would mainly introduce phase shifts in the oscillation
(\cf Press~1981).
For future use, we introduce the vertical and horizontal wavenumbers, $k_v$ and $k_h$,
\beq
k_v^2 =  \lp \frac{N^2}{\sigma^2} -1 \rp  \, \frac{\ell \lp \ell+1 \rp }{r^2}
= \lp \frac{N^2}{\sigma^2} -1 \rp  \, k_h^2.
\label{wavenb}
\eeq
This relation shows the highly anisotropic nature of internal waves, especially in the
stellar context where we are mainly looking at waves with $\sigma \ll N$.
When the wave is propagating, $\ell$ is conserved, $k_h$ varies as $1/r^2$
and $k_v$
is modified according to (\ref{wavenb}) when the local frequency 
$\sigma$ and/or the
Brunt-V\"ais\"al\"a frequency $N$ change. If $\sigma \ge N$, $k_v$ becomes imaginary and
the wave is reflected. 
When $\sigma \ll N$, we have $r k_v \gg 1$ and therefore  the
differential equation (\ref{secorder}) may be solved by the WKB method,
which yields 
\beq
u_r = C \; r^{-{3\over2}} \rho^{-{1 \over 2}}
\lp \frac{N^2}{\sigma^2} -1\rp^{-{1 \over 4}}
 P_\ell^m(\cos \theta)  
\cos \lp \sigma t - m \phi  -
\int_r^{r_c} k_v \diff r \rp, \label{wkbr}
\eeq
with $r_c$ designating the base of the convection zone\footnote{Here, we are
implicitly assuming that IGWs are generated in the outer convection zone and 
travel inwards.} and
$C$ is related to the wave amplitude. It describes a
monochromatic wave of local frequency
$\sigma$, which propagates\footnote{The most general solution would include a
stationary wave.}
with a vertical group velocity
\beq
v_g = {\diff \sigma \over \diff k_v} = -{\sigma \over k_v} \;
{N^2 - \sigma^2 \over N^2}
\label{vgroup}
 \eeq
and horizontal velocities
\beq
u_\phi =  -m \, { r k_v\over \ell(\ell+1)}\; {u_r
\over \sin \theta}
\label{wkbphi}
\eeq
and
\beq
u_\theta^2 + u_\phi^2  =  \lp \frac{N^2}{\sigma^2} -1 \rp \, u_r^2 .
\label{wkbh}
\eeq
The $\phi$ component of the velocity field depends on the azimuthal number $m$. In a
rotating star, we distinguish between {\em prograde} ($m>0$) and {\em retrograde} 
($m<0$) waves, corresponding respectively to waves traveling in the direction 
of rotation or against it\footnote{Note that some authors, including Unno et al.~(1989),
define the wave velocity as
$u_r \propto \exp \lp i m\phi\rp $
instead of $u_r \propto \exp \lp -i m\phi\rp$ as used here.}.

So far, nonadiabatic effects have not been taken into account. 
To do so, we follow Press~(1981) once more, and consider damping by
thermal diffusion in the Boussinesq approximation\footnote{Here, the continuity equation
(\ref{cont}) is replaced by $\vec{\nabla }\cdot \vec{u}=0$, and implies that vertical
variations of density are not considered in that equation.} and in Cartesian
coordinates\footnote{This derivation with mean molecular weight gradients is taken from
Zahn et al.~(1997).}. In that case, the equation of motion reads 
\beq
i \sigma \nabla^2 (\rho u_r) = g k_h^2 \rho'
\label{motion1}
\eeq
while the equation of state may be written 
\beq
{\rho' \over \rho} = - \delta {T' \over T} + \varphi {\mu' \over \mu},
\label{etat}
\eeq
where we again ignored the pressure perturbation. 
The fluctuation of $\mu$ is given by the conservation of mean molecular weight 
in the wave motion
\beq
 i \sigma {\mu'\over \mu} = {\nabla_\mu \over H_P}\, u_r ,
 \label{consmu}
\eeq
and that of $T$ by the heat equation 
\beq
i \sigma {T' \over T} = -  \lp \nabla_{\rm ad} - \nabla \rp \, {u_r\over H_P} +
{K \over T} \nabla^2 T' , \label{heat}
\eeq
where $K$ is the thermal diffusivity.
Combining Eqs.~(\ref{etat}), (\ref{consmu}) and (\ref{heat}), we obtain
\beq
\lp i \sigma - K \nabla^2 \rp \rho' = \frac{N^2}{g} \rho u_r 
- \frac{K}{i \sigma} \nabla^2 \lp {N_\mu^2 \over g} \rho u_r \rp,
\label{rhop}
\eeq
and with (\ref{motion1}), we get the wave equation
\beq
\left( \nabla^2 + k_h^2 {N^2 \over \sigma^2} \right) \rho u_r
+ i {K \over \sigma}  \nabla^2
\left( \nabla^2 + k_h^2 {N_\mu^2 \over \sigma^2} \right) \rho u_r = 0 . 
\label{waveeq2}
\eeq
From this equation, we derive the
dispersion relation
\beq
\lc k_v^2 -   k_h^2 \lp \frac{N^2}{\sigma^2} - 1 \rp
\rc  -  i \, \frac{K}{\sigma} \lc k_v^2 + k_h^2 \rc
\lc k_v^2 - k_h^2 \lp \frac{N_\mu^2}{\sigma^2} - 1 \rp \rc = 0 .
\label{reldisp}
\eeq

We now have Eq.~(\ref{secorder}) which is the adiabatic equation in spherical
geometry while Eq.~(\ref{waveeq2}) takes into account non-adiabaticity.
We combine wanted features from both these equations to obtain an equation
that is locally Boussinesq and globally anelastic, and that takes into account
nonadiabatic effects
\beq
\frac{\diff^2 \Psi}{\diff r^2} + \lp \frac{N^2}{\sigma^2} -1 \rp
{\ell(\ell+1)\over r^2} \Psi + i \frac{K}{\sigma}  \nabla^2
\lp \nabla^2 + k_h^2 \frac{N_\mu^2}{\sigma^2} \rp \Psi = 0 \,.
\label{waveeq}
\eeq
As long as thermal damping is not too large\footnote{If the imaginary part of this equation is
larger than the real part, the solution for $u_r$ is evanescent rather than
oscillatory.}, the WKB solution for this equation is still given by equation~(\ref{wkbr}),
but multiplied by a damping factor $\exp (-\tau/2)$ where
\beq
\tau(r,\sigma,\ell) = \lc \ell(\ell+1) \rc ^{3 \over 2} \int_r^{r_c} K  \; {N \, N_T^2 \over
\sigma^4}  \lp \frac{N^2}{N^2 - \sigma^2}\rp^{1 \over 2} \frac{\diff r}{r^3}.
\label{tau}  
\eeq


\subsection{Excitation of internal waves \label{sec:exc}}

In the Earth's atmosphere, IGWs are generated by several processes, 
like \eg
conv\-ection and wind compression by topography. 
In the single star context, we are interested by waves
produced by the injection of kinetic
energy from a turbulent region to an adjacent stable region\footnote{In binary stars,
the variation of the gravitational potential can excite a gravity wave in early type
stars. Its effect has been studied by Zahn~(1975a) and Goldreich \& Nicholson~(1989).}.
In the Earth's atmosphere, this is observed at the border of clouds 
(Townsend~1965). 
The existence of those waves in stars is observed in 
numerical simulations of penetrative convection both in 2--D and 3--D 
(Hurlburt et al.~1986,~1994, 
Andersen~1994, Nordlund et al.~1996, Kiraga et al.~2000, Dintrans et al.~2005, Rogers \& 
Glatzmeier~2005a,b). 
They are excited by two different processes namely \vspace*{-0.2cm}
\begin{itemize}
\item convective overshooting in a stable region;\vspace*{-0.2cm}
\item excitation by the Reynolds stresses in the convection zone
itself. \vspace*{-0.17cm}
\end{itemize}
Both sources contribute to the excitation.

Ultimately, one would like to obtain realistic wave fluxes from numerical simulation
themselves. Some improvement for such prescriptions has been made by
Rogers \& Glatzmeier~(2005b) who calculated wave excitation in a cylindrical (2--D) 
model having a 
stratification that is similar to that of the Sun. However, the level of
turbulence reached in that simulation is far from realistic and thus, cannot provide a
good answer to the question yet. Furthermore, an analysis of the wave spectrum
characteristics as a function of convective properties (such as the convective
luminosity and/or the turn-over time-scale) has not been done. For stellar applications,
the variation of wave generation with effective temperature\footnote{One should note 
that the
properties of the surface convection zone, like its depth and relevant
time-scales, are determined 
almost uniquely by the star's surface temperature.} is a key feature in understanding
their differential properties (see \S\,\ref{sec:li}). These are the two reasons why we will rather rely on
theoretical estimates for wave generation.

Excitation by overshooting is quite difficult to evaluate analytically. 
Garc\'\i a L\'opez \& Spruit~(1991) made such an attempt and assumed that the
pressure perturbation produced by turbulent eddies
at the radiative/convective boundary is equal to the wave pressure perturbation.
In this model, small stochastic eddies combine 
to excite a whole wavelength spectrum. It has been
formulated under the assumption of homogeneous turbulence, but it could be
modified to incorporate the presence of down-drafts as observed in numerical
simulations. Comparisons of numerical results with this theoretical model show that
the maximum in the wave spectrum produced by the simulation was similar in
amplitude to that of this parametric model (Kiraga et al.~2003, Rogers \& Glatzmeier~2005b). 
However, in simulations, modes were excited over a
much broader range of frequencies and wavelengths. It is not clear whether this
is related to the bi-dimensional nature of the simulations or to shortcomings in 
the model. 

Another estimate for excitation by overshooting has been made by
Fritts, Vadas, \& Andreassen~(1998), in a study aimed at estimating the residual
circulation induced by latitude dependent wave dissipation in the tachocline. 
However, they concentrate on the small wavelength waves that dissipate 
close to the convection zone, and their mechanism does not take the combination
of small scale eddies into account in order to produce low degree waves. 
Such waves are essential in order to influence the inner regions on an evolutionary 
time-scale (\S\,\ref{sec:mom}, see also Talon, Kumar, \& Zahn~2002).
Let us also note that such low degree waves are actually observed in the wave spectra of
numerical simulations. 

IGWs can also be excited in the convection zone itself.
That can be estimated by convolving the local wave function, which is
evanescent and whose local amplitude is proportional to 
$\exp \lc -\int \diff r \, k_r \rc$, with entropy fluctuations and Reynolds stresses.
A complete description of this process has been developed by Goldreich and his
collaborators (Goldreich \& Keeley~1977, 
Goldreich \& Kumar~1990, Goldreich, Murray, \& Kumar~1994, hereafter GMK). Their model 
was first applied to the solar p-modes, and quite
successfully reproduces the solar spectral energy input rate distribution, 
provided one free parameter which describes the geometry of turbulent eddies is
calibrated. In that case, driving is dominated by entropy fluctuations.
Balmforth~(1992) made a similar study, using a somewhat different formalism. 
Subject to the calibration of a free parameter, he is also able to reproduce the spectral
energy distribution; however, Reynolds-stresses dominate the
driving in his calculations.

\begin{figure}
\begin{center}
  \includegraphics[width=5.5cm,angle=-90]{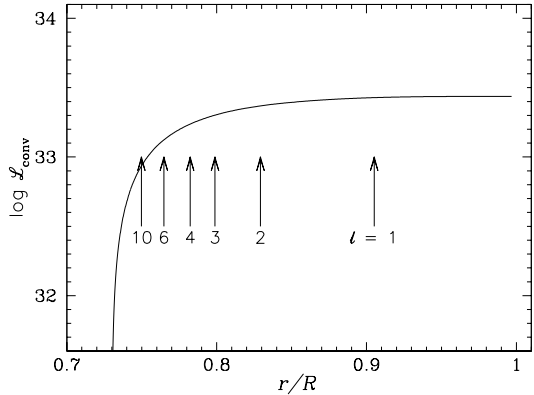}
  \vspace*{-0.5cm}
\end{center}
  \caption{Penetration of IGW into the convection zone of a ZAMS $1\,M_\odot$
model for various degrees $\ell$. The 
arrows indicate the 
depth where the amplitude is reduced by a factor of 2. 
Also shown is the local convective flux given by the mixing length theory.
\label{excitation}}
\end{figure}

The GMK formalism can be adapted to the case of g-modes (Kumar \& Quataert 1997);
the energy flux per unit frequency ${\cal F}_E$ is then
\begin{eqnarray}
{\cal F}_E \lp \ell, \sigma \rp &=& \frac{\sigma^2}{4\pi} \int \diff r\; \frac{\rho^2}{r^2}
   \left[\left(\frac{\partial \xi_r}{\partial r}\right)^2 +
   \ell(\ell+1)\left(\frac{\partial \xi_h}{\partial r}\right)^2 \right]  \nonumber \\
 && \times  \exp\left[ -h_\sigma^2 \frac{\ell(\ell+1)}{2r^2}\right] \frac{v^3 L^4 }{1
  + (\sigma \tau_L)^\frac{15}{2}},
\label{gold}
\end{eqnarray}
where
$\xi_r$ and $[\ell(\ell+1)]^{1 \over 2}\xi_h$ are the radial and horizontal
displacement wave-functions, which are normalized to unit energy flux just
below the convection zone, $v$ is the convective velocity, $L$ the radial
size of an energy bearing turbulent eddy, $\tau_L \approx L/v$ the
characteristic convective time, and $h_\sigma$ is the
radial size of the largest eddy at depth $r$ with characteristic frequency of
$\sigma$ or greater ($h_\sigma = L \min\{1, (2\sigma\tau_L)^{-\frac{3}{2}}\}$).

In the convection zone, the mode is
evanescent and the penetration depth\footnote{This is
the theoretical dependence. Damping can be enhanced by turbulence, which would reduce
the surface amplitude of the mode (see \eg Andersen~1996).} 
is inversely proportional to
$\sqrt{\ell \lp \ell +1 \rp}$.
Figure~\ref{excitation} compares this distance for various modes with the 
convective energy that is
locally available. In the outer part of the model, the luminosity is carried almost 
uniquely by convection.

\subsection{Particle transport\label{sec:part}}

While we all learned in basic physics lectures that waves do not produce 
transport of matter, this is actually only true when a first order, linear 
approximation is used. In fact, there exists a wide range of physical 
processes that may give rise to mixing by
waves, and we will briefly review some of these here. Let us note however that the 
actual calculations that were made with these in a stellar context do not 
take into account angular momentum transport by the same waves,
which would greatly reduce their penetration depth 
(see \S\,\ref{sec:mom}) and thus, the (numerical) results that are
mentioned should be reviewed 
accordingly.


\subsubsection{Shear unstable waves \label{sec:shearwave}}

Let us look at the case of low frequency IGWs in which $\sigma \ll N$. Then, 
the velocity of the associated flow is nearly horizontal, and produces a 
sinusoidally varying vertical shear. Such a flow has inflection points, which 
may trigger a shear instability that will produce small scale turbulence. 
Because the flow is time-dependent, the shear rate $k_v u_h$ has to be larger 
than the wave frequency 
for this to occur and hence the condition for weak mixing by this 
process is (Press~1981, Garc\'\i a L\'opez \& Spruit~1991)
\beq
k_v u_h > \sigma.
\eeq
When this condition is satisfied, the local shear rate can be compared to 
the stabilizing Brunt-V\"ais\"al\"a frequency, leading to the well known 
Richardson criterion. Instability occurs when
\beq
Ri = \frac{N^2}{\lp k_v u_h \rp^2} < \frac{1}{4}.
\eeq
Taking thermal diffusivity into account, 
the Richardson criterion is modified and becomes
\beq
\frac{N^2}{\lp k_v u_h \rp ^2} \, \frac{\ell_0 v_0}{K} < \frac{1}{4}
\eeq
where $\ell_0 v_0 = D_0$ is the turbulent diffusion associated with eddies of 
size $\ell_0$ and velocity $v_0$. The actual turbulent
diffusion $D_{\rm wave\, shear}$
will be given by the largest value of this product that is unstable and thus
\beq
D_{\rm wave\, shear} = \frac{1}{4} \, \frac{\lp k_v u_h \rp ^2}{N^2}\, K.
\eeq

Garc\'\i a L\'opez \& Spruit~(1991) applied such a formula to the case of 
F stars, in order to reproduce the so-called Li dip (see \S\,\ref{sec:li}). 
Using an excitation model based on the matching of pressure fluctuations at 
the base of the convection zone and that of the waves, they showed that Li 
burning could be produced in the dip region
if the input of turbulent energy in waves is large enough\footnote{In their paper,
Garc\'\i a L\'opez \& Spruit multiplied the wave flux calculated using
a standard mixing length approach of convection
by 10 to obtain enough depletion in the Li-dip. This 
could be explained by considering the presence of plumes in convection.} and provided the 
mixing length of convection is suitably chosen.

\begin{figure}
\begin{center}
  \includegraphics[width=10cm]{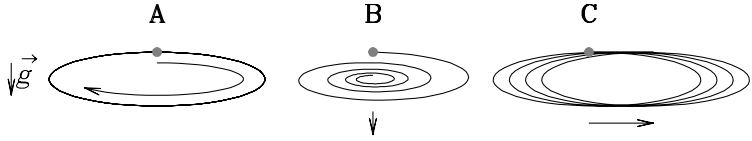}
  \vspace*{-0.5cm}
\end{center}
  \caption{Movement of a fluid parcel (whose initial position is shown with a gray
  dot) when a wave passes in the linear
  regime ({\bf A}), when thermal diffusion is taken into account ({\bf B}) and when
  non-linearities are taken into account ({\bf C}). In ({\bf B}), the net 
  displacement is in
  the direction of gravity while in ({\bf C}), it is perpendicular to $\vec{g}$.
\label{stokes}}
\end{figure}

\subsubsection{Irreversible second-order stochastic diffusion}

In the adiabatic and linear limit, the trajectory of a fluid parcel as a wave passes is 
a simple circular movement (Fig.~\ref{stokes}{\bf A}). However, when thermal diffusion 
is taken into account, the ``memory'' of the initial entropy content is lost, and 
the fluid parcel does not return to its initial entropy level 
(Fig.~\ref{stokes}{\bf B}; Press~1981, Schatzman~1993, Montalb\'an~1994). 
In the adiabatic limit, the vertical displacement of
the fluid element is given by $\delta_v = u_v/\sigma$, while in the absence of 
a restoring
force it would be $\Delta_v=k_v \delta_v ^2$. The fraction of entropy that is lost 
during a vertical
excursion can be roughly estimated as the fraction of this length-scale $\Delta_v$
corresponding to
the ratio of the diffusive time-scale $K \, k_v^2$ and the period $1/\sigma$. 
Overall, the vertical displacement of the fluid parcel is
\beq
\ell = K \frac{k_v^3\,u_v^2}{\sigma^3}
\eeq
(Press~1981).
The velocity scale associated with this process is
\beq
v = \ell \sigma,
\eeq
and this leads to a diffusion coefficient of the form
\beq
D_{\rm 2^{\rm nd}~order} \equiv \ell v = K^2 \, \frac{k_v^6\, u_v^4}{\sigma ^5}
\eeq
for one wave. A combination of all excited waves has to be made to transform 
this into a
global diffusion coefficient (this is discussed at length by Montalb\'an~1994).
This
process should be efficient only over about one damping length scale and thus, 
will lead 
to a diffusion coefficient that decreases quite rapidly with depth.
Calculations using such a description were made by Montalb\'an \& Schatzman~(1996)
to compute Li destruction in cool main sequence stars.

\subsubsection{Stokes' drift}

{\em Stokes' drift} can be defined as the difference between the mean 
fluid velocity measured 
at a point and the mean velocity of a drifter in the direction of wave propagation
and was originally described by
Stokes (1847, see Lighthill~1979 for a complete description). 
If there is no mean velocity, it represents
the velocity of a passive scalar that is slowly moving through the fluid. This displacement
is in the direction of wave propagation ($\vec{v_p}$), which is mainly horizontal for the
IGWs we are studying here. Heuristically, this drift is caused by an asymmetry between the 
horizontal displacement when the wave is moving downward and upward. The resulting 
movement is shown in Fig.~\ref{stokes}{\bf C}. In the case of stars, this will
produce mainly a horizontal redistribution of elements and thus, little
vertical mixing is expected from this process.

\subsubsection{Wave-induced turbulence}

The last process I will describe here is wave-induced turbulence.
This has been discussed by a few authors, under various forms.
Canuto~(2002) developed a model in which gravity waves
act as a source term in the equation that describes turbulence.
Young et al.~(2003) describe mixing by waves generated at the boundary
of a convective region as a physical description of ``overshooting''.
Talon \& Charbonnel~(2005) study the formation of an oscillating shear layer
caused by differential wave damping (see \S\,\ref{sec:SLO}). In that process,
energy is transfered 
by waves from the convection zone to the shear layer and stored in 
differential rotation, which can then be converted to turbulence by the shear
instability (\S~\ref{sec:shear}). Let us note that it is different from
the process described in \S~\ref{sec:shearwave}: here it is the rotation
profile that becomes unstable, and not the wave itself.

In all these cases, turbulent mixing should be large in a rather small region
just below the convection zone, and acts to smooth out the possible 
chemical discontinuity at convective boundaries.

\subsection{Momentum transport\label{sec:mom}}

Let us now describe how angular momentum is transported by IGWs\footnote{This
description follows Zahn et al.~(1997).}.
First, the horizontal average of the kinetic energy density can be calculated
from the WKB velocities (\ref{wkbr}) and (\ref{wkbh}) 
and yield
\beqan
{1\over2}\, \rho <u^2> &\equiv&  {1 \over 4 \pi} \int\!\!\!\int  {1\over2} 
\; \rho (u_r^2 + u_\theta^2 + u_\phi^2) \, \sin \theta \; \diff \theta \,
\diff \phi  \nonumber  \\ 
&=& {1\over2}\; {N^2 \over \sigma^2} 
{1 \over 4 \pi} \int\!\!\!\int \rho \; u_r^2 \, \sin \theta \; \diff \theta 
\, \diff \phi \equiv  {1\over2}\, {N^2 \over \sigma^2}\,
\rho\, <u_r^2>. 
\label{endensity}
\eeqan
We now multiply this quantity by the group velocity (\ref{vgroup}) to
get the average kinetic energy flux carried by a traveling IGW
\beq
{\cal F}_{K,\sigma,\ell,m} = -{1\over2} \, \rho <u^2> \,{\sigma \over k_v} 
{N^2 - \sigma^2 \over N^2} 
 =  -{1\over2} \, \rho <u^2> \, {\sigma^2 \over N^2} 
\, {(N^2 - \sigma^2)^{1 \over 2} \over k_h}.
\label{fluxk}
\eeq

\begin{figure}
\begin{center}
  \vspace*{-1cm}
  \includegraphics[width=7.2cm]{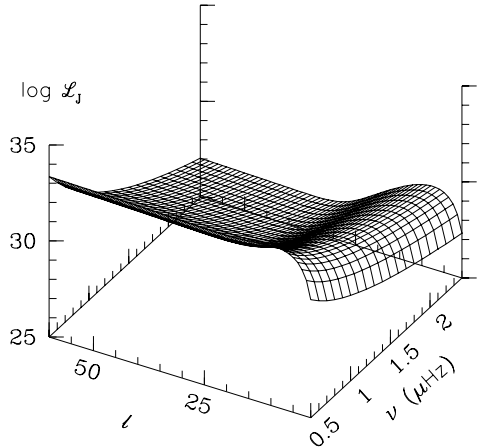}
  \vspace*{-1cm}
\end{center}
  \caption{Angular momentum wave luminosity ${\cal L}_{J,\sigma,\ell}$ in a ZAMS $1\,M_\odot$
model for various degrees $\ell$ and frequencies $\nu=\sigma/2\pi$ according to the GMK model.
\label{spectre}}
\end{figure}

The mean radial flux of angular momentum is evaluated using Eqs.~(\ref{wkbr}), 
(\ref{wkbphi}), (\ref{endensity}) and (\ref{fluxk})
\beqan
{\cal F}_{J,\sigma,\ell,m} &=& {1 \over 4 \pi} \int\!\!\!\int   \rho\, r \sin \theta\, 
u_\phi u_r \,  \sin \theta \; \diff \theta \, \diff \phi 
\nonumber  \\
&=&  -m \,r \,{r k_v \over \ell(\ell+1)}{1 \over 4 \pi} \int\!\!\!\int 
\rho u_r^2 \, \sin \theta \; \diff \theta \, \diff \phi  
= 2 \,{m \over \sigma} \; {\cal F}_{K,\sigma,\ell,m} .
\label{fluxj}
\eeqan
Inserting (\ref{wkbr}) into (\ref{fluxj}), it is possible to
verify that, in the adiabatic limit, the angular momentum luminosity 
is conserved\footnote{On the other hand, the kinetic
energy of the wave varies with depth since the local frequency $\sigma$ 
depends on the rotation rate $\Omega(r)$.
Bretherton~(1969) showed that, in the plane-parallel case, conservation is
ensured in the inertial frame.}
\beq
4 \pi r^2 {\cal F}_{J,\sigma,\ell,m}(r) \equiv {\cal L}_{J,\sigma,\ell,m} = \hbox{cst}.
\eeq
Figure~\ref{spectre} shows the wave spectrum of angular momentum
luminosity in a $1~M_\odot$ model when the GMK
model is used for the kinetic energy flux.

When radiative damping is taken into account, 
the wave amplitude is multiplied by an attenuation factor $\exp
(-\tau/2)$ (\cf Eq.~\ref{tau}).
Locally, the total angular momentum luminosity is the sum of the contribution
of all waves
\beq
{\cal L}_J(r) = \sum_{\rm waves} {\cal L}_{J,\sigma,\ell,m}(r_c)\,  
\exp \lc -\tau(r,\sigma,\ell) \rc.
\label{dampj}
\eeq
The deposition of angular momentum is then given by the radial derivative
of this luminosity. Let us remark that here we consider only the radial dependency
of wave transport, thus the use of horizontal averaging.
When other processes for angular momentum redistribution are also taken into
account, 
the evolution of angular momentum follows
\beq
\rho \dtt \lc r^2 {\Omega}\rc = 
\frac{1}{5 r^2} \drr \lc \rho r^4 \Omega U \rc 
+ \frac{1}{ r^2} \drr \lc \rho \nu_t r^4 \dr{\Omega} \rc 
\pm \frac{3}{8\pi} \frac{1}{r^2} \drr{{\cal L}_J(r)},
\label{ev_omega}
\eeq
where $U$ is the radial meridional circulation velocity and
$\nu_t$ the viscosity from hydrodynamical instabilities (\S~\ref{sec:rot})\footnote{One 
could also add the magnetic viscosity of \S~\ref{sec:TS}, but for consistency 
gravito-Alfv\'en waves should then be considered. See \eg Barnes, MacGregor \& Charbonneau~1998.}. 
The ``$+$'' (``$-$'') sign in front of the angular 
momentum luminosity corresponds to a wave traveling inward (outward).

\begin{figure}
\begin{center}
    \includegraphics[width=5.5cm,angle=-90]{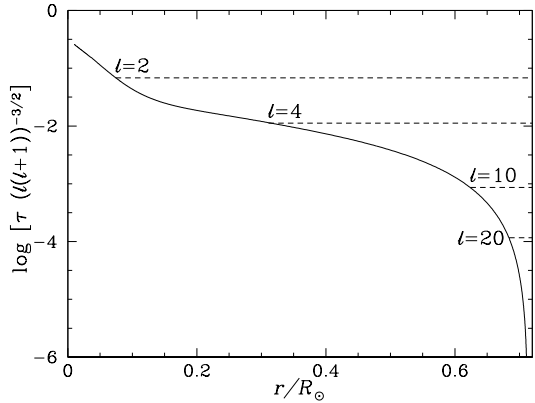}
  \vspace*{-0.5cm}
\end{center}
  \caption{Evaluation of the damping factor $\tau$ (\cf Eq.\ref{tau}) for a
  frequency of $1~\mu{\rm Hz}$ in a solar model. 
  The depth corresponding to an attenuation by a
  factor $1/e$ is shown for various degrees $\ell$.
\label{damping}}
\end{figure}

\subsubsection{Wave-mean flow interaction and the SLO \label{sec:SLO} }

We now examine the impact of IGWs on the radial distribution of
angular momentum. Figure~\ref{damping} shows the penetration depth for
waves of various degrees. For low frequency waves, damping
is strong enough so that most waves are damped very close to the base of the
convection zone. These are the waves we will examine here.

\begin{figure}
\begin{center}
    \includegraphics[width=10cm]{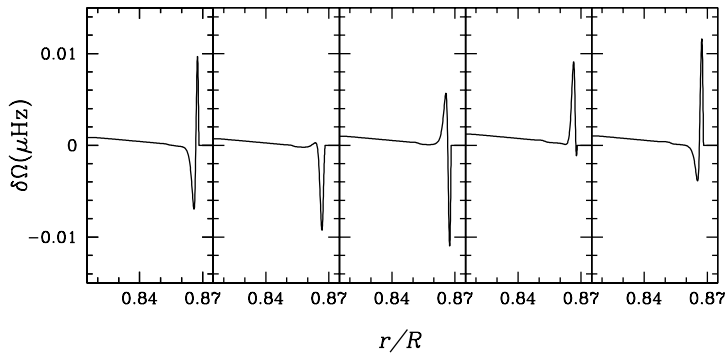}
    \vspace*{-0.5cm}
\end{center}
  \caption{Shear layer oscillation (SLO) in a $1.3~M_\odot$ model. Successive profiles
  are separated by 1 year intervals. The layer's thickness depends
  on the magnitude of thermal dissipation and the oscillation period is a function
  of the total wave angular momentum luminosity.
\label{SLO}}
\end{figure}

Let us first take a look at the damping integral (\ref{tau}), and assume that both
prograde and retrograde waves are excited with the same amplitude (according to
Eq.~\ref{gold}). In solid body rotation, both waves are equally dissipated 
when traveling inward and
there is no impact on the distribution of angular momentum. In the
presence of differential rotation, the situation is different. If the interior is
rotating faster than the convection zone, the local frequency of prograde waves
diminishes, which enhances their dissipation; the corresponding retrograde waves 
are then
dissipated further inside. This produces an increase of the local
differential rotation, and creates a double-peaked shear layer\footnote{Because
local shears are amplified by waves, even a small perturbation will trigger this.}. 
In the presence of
shear turbulence, this layer oscillates, producing a ``shear layer oscillation''
or SLO
(\cf Fig.~\ref{SLO} and Gough \& McIntyre~1998, Ringot~1998, 
Kumar, Talon, \& Zahn~1999). Viscosity is essential
in this process and is responsible for the disappearance of the outermost peak as its
slope gets steeper and the peak is slowly absorbed by the convection zone.
This oscillation is similar to the Earth's QBO (see \eg McIntyre~2003 for details regarding 
the QBO).

\subsubsection{Secular effects}

The SLO, by locally Doppler shifting to lower frequencies both prograde and retrograde
waves, acts as an efficient IGW filter (see Eq.~\ref{tau}) and for some
time there was doubt that any wave power would remain beyond 
(Gough \& McIntyre~1998). 
However, Talon, Kumar, \& Zahn~(2002) and Talon \& Charbonnel~(2005) performed 
complete calculations of wave dissipation in the presence of a SLO
and showed that
the low-degree waves do conserve a significant amplitude beyond the shear layer.

If the radiative zone has the same rotation rate as the convection
zone, over a complete SLO cycle, the
magnitude of the prograde and retrograde peaks\footnote{The prograde peak refers to
the layer that is rotating more rapidly than the convection zone and it produces
filtering of prograde waves, while the retrograde peak is rotating slower and
filters retrograde waves.} are on average equal,
and waves of azimuthal number $+m$ and $-m$
have (on average) equal amplitudes after crossing the SLO. In the presence of differential rotation
however, with the inner part rotating faster than the outer part as is the case of
a solar type star that is spun down by magnetic torque, the prograde peak is
always larger than the retrograde peak and there is a net flux of negative angular
momentum to be redistributed in the star's radiative zone. Talon,
Kumar, \& Zahn~(2002) showed that this can spin down the interior of a star
over long time-scales. Talon \& Charbonnel~(2005) later showed 
that this {\em asymmetric
filtering} depends only on the difference of rotation rates at the base of the
convection zone and at the base of the SLO 
$(\delta \Omega= \Omega_{\rm cz}-\Omega_{\rm SLO})$.

\begin{figure}
\begin{center}
    \hspace*{-0.25cm}\includegraphics[width=13cm]{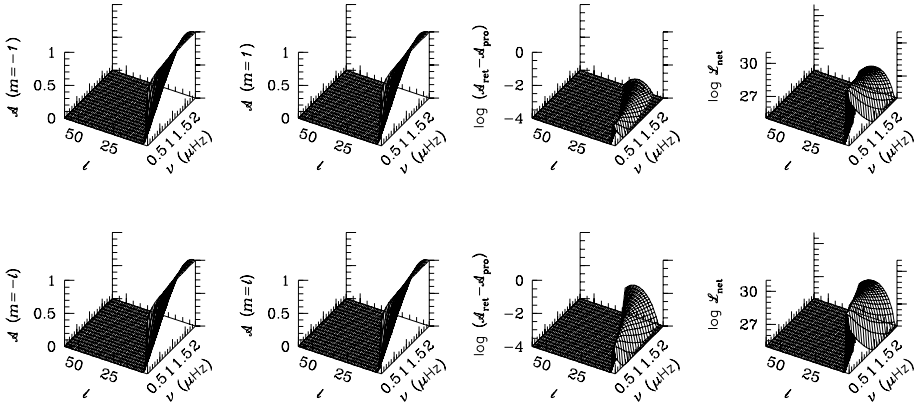}
    \vspace*{-0.5cm}
\end{center}
  \caption{Wave characteristics below the shear layer for a differential rotation
of $\delta \Omega = 0.1~\mu {\rm Hz}$ over $5\% $ in radius in a $1.3~M_\odot$ star.
{\em First and second columns}: Wave amplitude for $m=\pm 1$ ({\em top)}
and $m=\pm \ell$ ({\em bottom}).
{\em Third column}: Amplitude difference between retrograde and prograde waves of
the same azimuthal number.
{\em Fourth column}: Net luminosity for a given azimuthal number.
\label{filter}}
\end{figure}

Figure~\ref{filter} shows how wave amplitudes are differentially 
reduced by crossing the shear layer. The rightmost column shows the
net luminosity beyond the SLO; the low-frequency, low-degree
waves have the strongest impact on the secular redistribution of angular
momentum in the radiative part of the star. They must be low-frequency,
otherwise differential damping is too weak, and low-degree, otherwise they are
damped in the SLO rather than beyond.

\begin{figure}
\begin{center}
    \includegraphics[width=5.5cm,angle=-90]{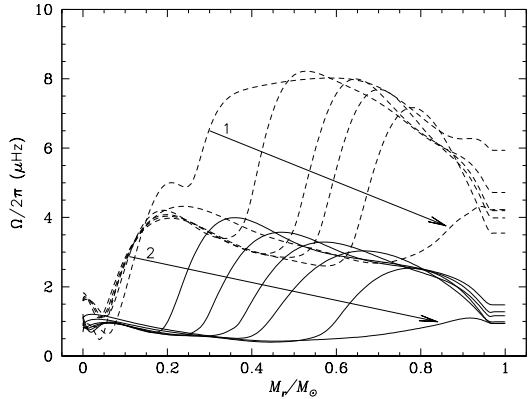}
  \vspace*{-0.5cm}
\end{center}
  \caption{Evolution of the interior rotation profile in a $1\,M_\odot$ model with
  meridional circulation, shear instabilities and IGWs. The initial equatorial rotation velocity is $50~{\rm km \,s^{-1}}$ 
  and surface magnetic braking is applied. Curves correspond to ages of 
  0.2, 0.21, 0.22, 0.23, 0.25,0.27 (dashed lines), 0.5, 0.7, 1.0, 1.5, 3.0 and 
  4.6 (solid lines) Gyr. {\em Adapted from Charbonnel \& Talon~(2005).}
\label{sunrot}}
\end{figure}

\subsection{Applications of angular momentum transport \label{sec:app}}

\subsubsection{The solar rotation\label{sec:sun}}

This description of IGWs has been applied to the solar rotation problem.
Talon, Kumar, \& Zahn~(2002) proved, in a static solar model, that IGWs 
could be responsible for the required angular momentum extraction. More
recently, Charbonnel \& Talon~(2005) revisited the problem and looked at the role
of IGWs in an evolving solar model, which undergoes spin-down from the
zero age main sequence (ZAMS) on. They showed that, at the age of the Sun,
differential rotation indeed becomes quite small (see Fig.~\ref{sunrot}).
In this simulation, the low-degree waves penetrate all the way to the core and 
spin it down extremely efficiently at the very beginning of the evolution. 
This is due to the small amount of angular momentum ($\propto r^2$) contained in the 
core. Once the core has been spun down, the damping of retrograde waves, that 
carry the negative angular momentum, increases locally. Consequently a 
``slowness'' front forms and propagates in a wave-like way from the core 
to the surface. As further braking takes place, a second front forms and 
propagates outward. The differential wave 
filtering adjusts itself so as to compensate for
the flux of angular momentum that is 
lost through the stellar wind. This explains why front propagation is fast at 
the beginning and then slows down, just as the spin-down rate does.

\begin{figure}
\begin{center}
    \includegraphics[width=5.5cm,angle=-90]{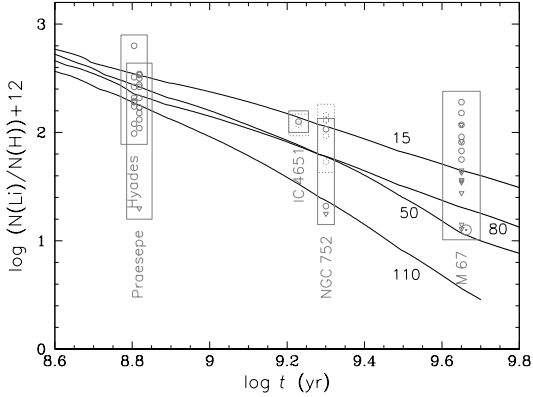}
  \vspace*{-0.5cm}
\end{center}
  \caption{Evolution of the surface lithium abundance for various initial rotation
  velocities in models including meridional circulation, shear instabilities and IGWs. 
  The surface Li abundances as measured in
  various clusters are also shown. The vertical extent of boxes shows the range 
  of lithium values for stars with an effective temperature corresponding to 
  that of the model $\pm 100 ~{\rm K}$ at the cluster age plus a typical error in 
  abundance determination. {\em Adapted from Charbonnel \& Talon~(2005).}
\label{lisun}}
\end{figure}

Figure~\ref{lisun} shows the evolution of the
surface Li abundance in a solar model for various initial velocities. One should
note here the (small) dispersion that is obtained when very
different initial velocities are considered. This would not be the case if only
meridional circulation and shear turbulence where considered since then, mixing
would be proportional to the angular momentum loss.

\subsubsection{The Li dip\label{sec:li}}

\begin{figure}
\begin{center}
    \includegraphics[width=5.5cm,angle=-90]{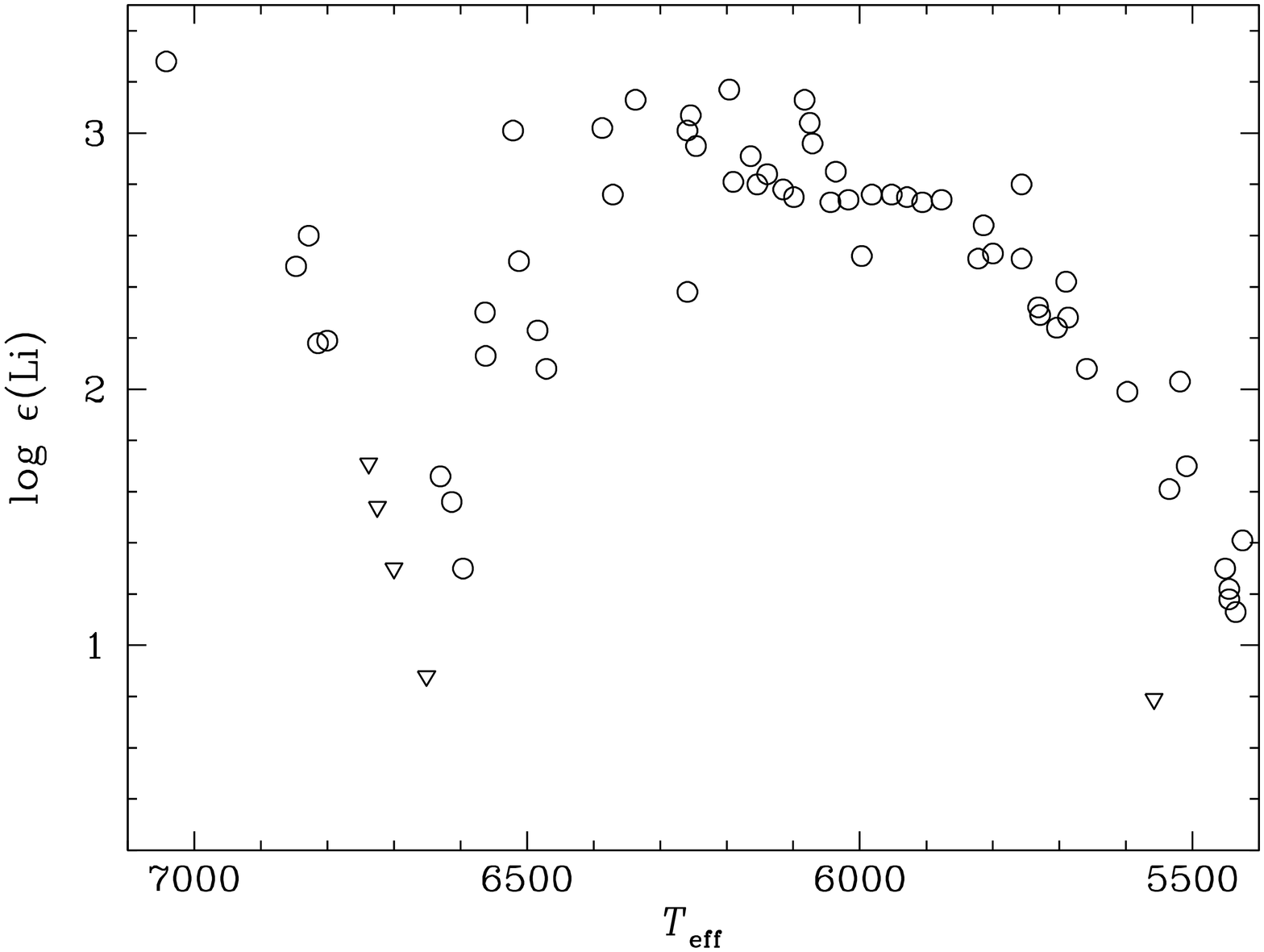}
    \includegraphics[width=5.5cm,angle=-90]{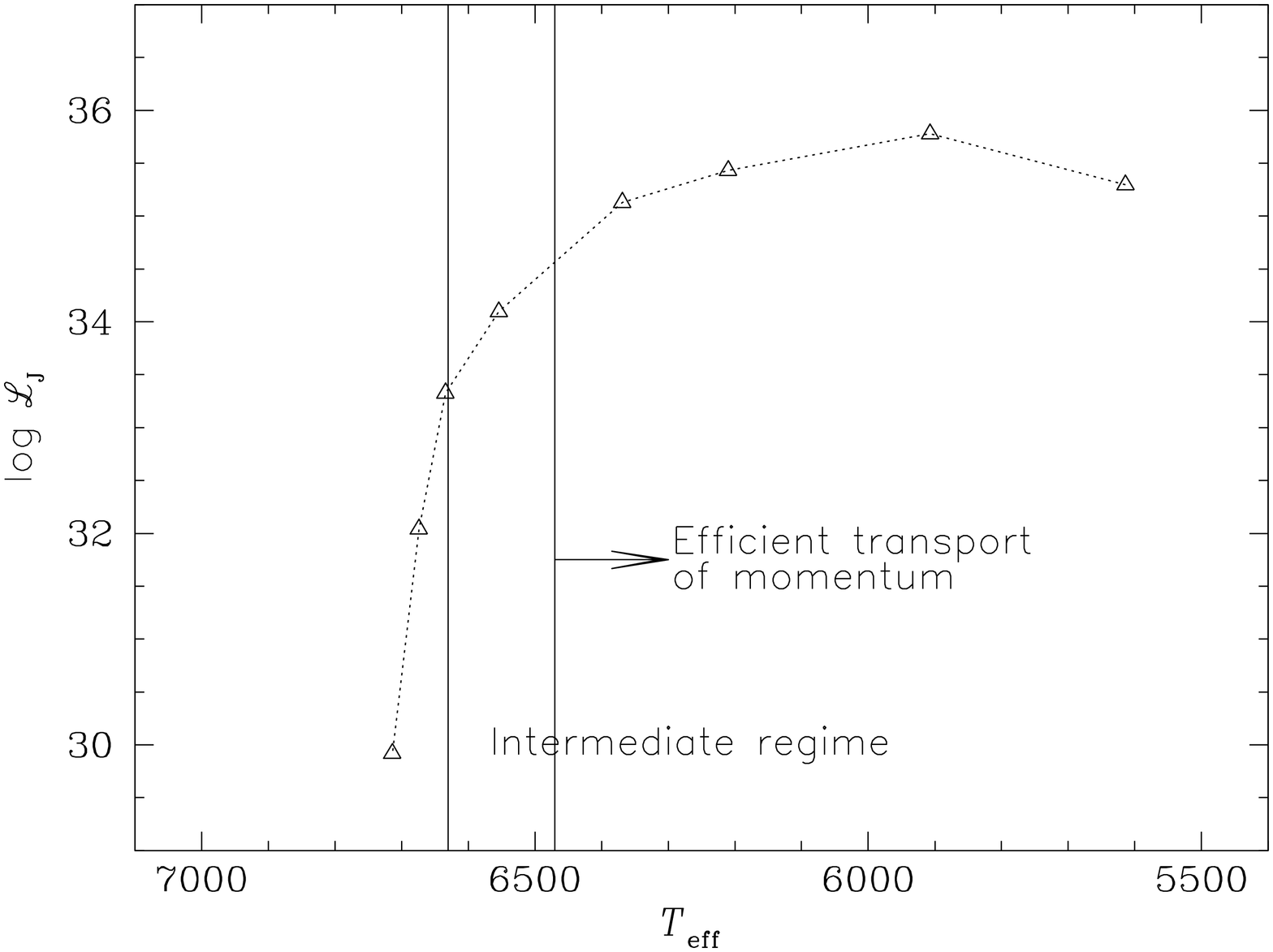}
  \vspace*{-0.5cm}
\end{center}
  \caption{({\em top}) Lithium dip as observed in the Hyades.
  ({\em bottom}) Net angular momentum wave luminosity below the SLO for a differential rotation
  $\delta \Omega = 0.05~\mu{\rm Hz}$ over $0.05~R_*$. The vertical lines
  correspond to the requirements in the efficiency of angular momentum 
  transport by a ``non-standard'' process so that rotational
  mixing would cause the Li dip as suggested by Talon \& Charbonnel~(1998). 
  {\em Adapted from Talon \& Charbonnel~2003.}
\label{Lidip}}
\end{figure}

We mentioned (\S~\ref{sec:approt}) that rotational mixing could explain the hot
side of the Li dip but destroy too much lithium when looking at cooler stars.
TC98 suggested that if the process that is required to explain the Sun's rotation
profile
starts to become efficient in the center of the Li dip, these observations would be
reconciled with rotational mixing, thus giving a model that is valid for {\em all} main
sequence stars.
One then identifies three regimes: \vspace*{-0.15cm}
\begin{itemize}
\item Stars with $T_{\rm eff} \gta 6900~{\rm K}$ have 
a very shallow convective envelope which is not an efficient site for 
magnetic generation via a dynamo process. They are thus not slowed down 
by a magnetic torque. As a result, these stars soon reach 
the equilibrium meridional circulation (\S~\ref{sec:circumom}) and
the associated weak mixing is 
just sufficient to counteract atomic diffusion.
\vspace*{-0.15cm}
\item In stars with $\sim 6900 \gta T_{\rm eff} \gta 6600~{\rm K}$, 
the convective envelope deepens and 
a weak magnetic torque appears that slows down the outer layers. In this case
the transport of angular momentum by meridional circulation and
shear turbulence increases, 
leading to a larger destruction of Li.
\vspace*{-0.15cm}
\item Stars on the cool side of the Li dip ($T_{\rm eff} \lta 6600~{\rm K}$) 
have an even deeper convective 
envelope sustaining a very efficient dynamo which produces a strong magnetic 
torque that spins down the 
outer layers very efficiently. This outer convection zone is an efficient source
for IGW generation and wave-induced transport reduces the efficiency of rotational
mixing, thus producing a rise in Li abundances on the cold side of the Li dip.
\vspace*{-0.15cm}
\end{itemize}
Talon \& Charbonnel~(2003) showed that the net angular momentum luminosity corresponding to
IGWs does have the proper temperature dependency 
(Fig.~\ref{Lidip}) and thus could well explain the cool side of
the lithium dip.
Complete evolutionary calculations are underway to verify this proposition.

\subsection{Open problems in internal waves physics}
While results obtained including internal waves in evolutionary models
are quite promising, they remain crude, and several points are yet to be addressed.
First of all, one should revisit the excitation of IGWs.
Recent progress has been made in the description of the excitation of
solar p-modes by considering better models for convection (see \eg Belkacem et al.~2006).
Work is underway to adapt this description to IGWs (Belkacem et al.~in prep.)
Another important point is the inclusion of the Coriolis force in calculations.
This issue has been raised by several authors (Talon~1997, Lee \& Saio~1997, Townsend~2003
and Mathis~2005) and work is underway to correctly incorporate it in
descriptions of momentum transport in massive stars (Pantillon, Talon \& Charbonnel~in prep.)
Finally, all calculations that are presented here correspond to horizontal
averages. Implicitly, this assumes that horizontal transport is efficient in the
redistribution of angular momentum along isobars. This faces the ``Rossby scale problem'' already
mentioned in \S~\ref{sec:open}.

\section*{Acknowledgments}
I would like to thank Jacques Richer for his assistance in preparing
the discussion on atomic diffusion and
for his careful reading of this manuscript 
which lead to major improvements.


\begin{thebibliography}{99}

\bibitem{} Acheson D.J.,~1978, Phil. Trans. Roy. Soc. Lond. A289, 459
\bibitem{} Alfv\'en H.,~1942, Nature 150, 405
\bibitem{} Aller L.H., Chapman S.,~1960, ApJ 132, 461
\bibitem{} Andersen B.N., 1994, Solar Phys. 152, 241
\bibitem{} Andersen B.N., 1996, A\&A 312, 610
\bibitem{} Ando H.,~1986, A\&A 163, 97
\bibitem{} Balmforth N.J.,~1992, MNRAS 255, 639
\bibitem{} Barnes G., Charbonneau P., MacGregor K.B.,~1999, ApJ 511, 466
\bibitem{} Barnes G., MacGregor K.B., Charbonneau P.,~1998, ApJ 498, L169
\bibitem{} Belkacem K., Samadi R., Goupil M.J., Kupka F., Baudin F.,~2006, A\&A 460, 183
\bibitem{} Boesgaard A.M., Tripicco M.J.,~1986, ApJ 303, 624
\bibitem{} Braithwaite J., Spruit H.C.,~2004, Nature 431, 819
\bibitem{} Braithwaite J.,~2006, A\&A 449, 451
\bibitem{} Bretherton F.P., 1969, Quart. J. R. Met. Soc. 95, 213
\bibitem{} Brun A.S., Zahn J.-P.,~2006, A\&A 457, 665
\bibitem{} Burgers J.M.,~1969, Flow equations for composite gases, Academic
Press, New York
\bibitem{} Canuto V.M.,~1998, ApJ 505, L47
\bibitem{} Canuto V.M.,~2002, MNRAS 337, 713
\bibitem{} Chaboyer B., Demarque P., Pinsonneault M.H.,~1995, ApJ 441, 865
\bibitem{} Chaboyer B., Zahn J.-P.~1992, A\&A  253, 173
\bibitem{} Chandrasekhar S.,~1961, Hydrodynamic and Hydromagnetic Stability,
Oxford University Press
\bibitem{} Chaplin W.J., Christensen-Dalsgaard J., Elsworth Y.,
Howe R., Isaak G.R., Larsen R.M., New R., Schou J., Thompson M.J., Tomczyk S.,
1999, MNRAS 308, 405
\bibitem{} Chapman S., Cowling T.G.,~1970, The Mathematical Theory of
Non-Uniform Gases, Cambridge University Press, 3rd ed
\bibitem{} Charbonneau P.~1992, A\&A  259, 134
\bibitem{} Charbonneau P., MacGregor K.B.,~1993, ApJ 417, 762
\bibitem{} Charbonnel C., Talon S.,~2005, Science 309, 2189
\bibitem{} Christensen-Dalsgaard J., Proffitt C.R., Thompson M.J.,~1993, ApJ
403, 75
\bibitem{} Couvidat S., Garc\'\i a R.A., Turck-Chi\`eze S., Corbard T., 
Henney C.J., Jim\'enez-Reyes S.,~2003, ApJ 597, L77
\bibitem{} Cowling T.G.,~1941, MNRAS 101, 367
\bibitem{} Cowling T.G.,~1951, ApJ 114, 272
\bibitem{} Dintrans B., Brandenburg A., Nordlund \r{A}, Stein R.F.,~2005,
A\&A 438, 365
\bibitem{} Eddington A.S.,~1925, Obs. 48, 73
\bibitem{} Eddington A.S.,~1930, in ``The internal constitution of the stars''
\bibitem{} Eggenberger P., Maeder A., Meynet G.,~2005, A\&A 440, L9
\bibitem{} Endal A.S., Sofia S.,~1976, ApJ 210, 184
\bibitem{} Endal A.S., Sofia S.,~1978, ApJ 220, 279
\bibitem{} Ferraro V.C.A.,~1937, MNRAS, 97, 458
\bibitem{} Fricke K.J.,~1968, Z. Astrophys., 68, 317
\bibitem{} Fritts D.C., Vadas S.L., Andreassen \O ., 1998, A\&A 333, 343
\bibitem{} Garaud P.,~2002, MNRAS, 335, 707
\bibitem{} Garc\'{\i}a L\'opez R.J., Spruit H.C.,~1991, ApJ 377, 268
\bibitem{} Goldreich P., Keeley D.A.,~1977, ApJ 212, 243
\bibitem{} Goldreich P., Kumar P.,~1990, ApJ 363, 694
\bibitem{} Goldreich P., Murray N., Kumar, P.,~1994, ApJ 424, 466
\bibitem{} Goldreich P., Nicholson P.D.,~1989, ApJ 342, 1079
\bibitem{} Goldreich P., Schubert G.,~1967, ApJ 150, 571
\bibitem{} Gough D. O., McIntyre M.E.,~1998, Nature 394, 755
\bibitem{} Gratton L.~1945, Mem. Soc. Astron. Ital. 17, 5
\bibitem{} H\o iland E.,~1941, Avhandgliger Norske Videnskaps-Akademi i Oslo,
I, Math.-Naturv. Klasse 11, 1
\bibitem{} Hurlburt N.E., Toomre J., Massaguer J.M.,~1986, ApJ 311, 563
\bibitem{} Hurlburt N.E., Toomre J., Massaguer J.M., Zahn J.-P.,~1994, ApJ 421, 245
\bibitem{} Kawaler S.,~1988, ApJ 333, 236
\bibitem{} Kiraga M., R\'o\.zyczka M., Stepien K., Jahn K., Muthsam H., 2000, Acta Astronomica,
50, 93
\bibitem{} Kiraga M., Jahn K., Stepien K., Zahn J.-P., 2003, Acta Astronomica,
53, 321
\bibitem{} Knobloch E., Spruit H.C.~1982, A\&A  113, 261
\bibitem{} Knobloch E., Spruit H.C.~1983, A\&A  125, 59
\bibitem{} Kumar P., Quataert E.J.,~1997, ApJ 475, L143
\bibitem{} Kumar P., Talon S., Zahn J.-P.,~1999, ApJ  520, 859
\bibitem{} Langer N.,~1991, A\&A 243, 155
\bibitem{} Lee U.,~2006, MNRAS 365, 677
\bibitem{} Lee U., Saio H.,~1997, ApJ 491, 839
\bibitem{} Lighthill J.,~1979, Waves in Fluids, Cambridge University Press, Cambridge
\bibitem{} MacGregor K.B., Charbonneau P.,~1999, ApJ 519, 911
\bibitem{} McIntyre M.E.,~2003, in ``Perspectives in Fluid Dynamics: A Collective
Introduction to Current Research'', Eds. G.K. Batchelor, H.K. Moffat, M.G. Worster,
Cambridge Univ. Press, 557
\bibitem{} Maeder A.,~1995, A\&A 299, 84
\bibitem{} Maeder A.,~2003, A\&A, 399, 263
\bibitem{} Maeder A., Meynet G.,~2000, ARAA 38, 143
\bibitem{} Maeder A., Meynet G.,~2004, A\&A 422, 225
\bibitem{} Maeder A., Zahn J.-P.~1998, A\&A  334, 1000
\bibitem{} Matias J., Zahn J.-P.~1998, Sounding solar and stellar
interiors, IAU Symposium 181, Nice,  Eds. J. Provost \& F.X. Schmider
\bibitem{} Mathis S., 2005, PhD thesis, 
Universit\'e Paris VII
\bibitem{} Mathis S., Palacios A., Zahn J.-P.,~2004, A\&A 425, 243
\bibitem{} Mathis S., Zahn J.-P.,~2005, A\&A 440, 653
\bibitem{} Mestel L.,~1953, MNRAS 113, 716
\bibitem{} Mestel L., Weiss N.O.,~1987, MNRAS 226, 123 
\bibitem{} Michaud G.,~1993, Phys. Scripta T47, 143
\bibitem{} Michaud G., Charland Y., Vauclair S., Vauclair G.,~1976, ApJ 210, 447
\bibitem{} Montalb\'an J.,~1994, A\&A 281, 421
\bibitem{} Montalb\'an J., Schatzman E.,~1996, A\&A 305, 513
\bibitem{} Montalb\'an J., Schatzman E.,~2000, A\&A 354, 943
\bibitem{} Nordlund A., Stein R.F., Brandenburg A., 1996, Bull. Astron. Soc. of India
24, 261
\bibitem{} \"Opik E.J.~1951, MNRAS  111, 278
\bibitem{} Palacios A., Talon S., Charbonnel C., Forestini M.,~2003, A\&A, 399, 603
\bibitem{} Parker E.N.,~1960, ApJ 132, 821
\bibitem{} Pinsonneault M.H., Kawaler S.D., Sofia S., Demarque P.,~1989, ApJ 338, 424
\bibitem{} Pitts E., Tayler R.J.,~1986, MNRAS 216, 139
\bibitem{} Press W.H.,~1981, ApJ 245, 286
\bibitem{} Rayleigh Lord,~1880, Proc. London Math. Soc., 11, 57
\bibitem{} Rayleigh Lord,~1916, Proc. Roy. Soc. London A93, 148
\bibitem{} Richard D. \& Zahn J.-P.,~1999, A\&A, 347, 734
\bibitem{} Richer J., Michaud G., Rogers F., Iglesias C., Turcotte S., 
Leblanc F.,~1998, ApJ 492, 833
\bibitem{} Richer J., Michaud G., Turcotte S.,~2000, ApJ 529, 338
\bibitem{} Ringot O.,~1998, A\&A 335, 89
\bibitem{} Rogers T.M., Glatzmeier G.A.,~2005a, ApJ 620, 432
\bibitem{} Rogers T.M., Glatzmeier G.A.,~2005b, MNRAS 364, 1135
\bibitem{} Schatzman E.,~1962, An.Ap. 25, 18
\bibitem{} Schatzman E.,~1993, A\&A 279, 431
\bibitem{} Schatzman E., Baglin A.~1991, A\&A 249, 152
\bibitem{} Solberg H.~1936, Proc\`es-Verbaux Ass. M\'et\'eor., UGGI,
6$^{e}$ Assembl\'ee G\'en\'erale, Edinburgh, M\'em. et Disc. 2, 66
\bibitem{} Spruit H.C.,~1999, A\&A 349, 189
\bibitem{} Spruit H.C.,~2002, A\&A 381, 923
\bibitem{} Stokes G.G., 1847, Trans. Camb. Philos. Soc. 8, 441
\bibitem{} Sutherland, B.R., Linden, F.R.,~2002, Phys. Fluids 14, 721
\bibitem{} Sweet P.A.~1950, MNRAS  110, 548
\bibitem{} Talon S.,~1997, ``Hydrodynamique des \'etoiles en rotation''
\bibitem{} Talon S., Charbonnel C.,~1998, A\&A 335, 959
\bibitem{} Talon S., Charbonnel C.,~2003, A\&A 405, 1025
\bibitem{} Talon S., Charbonnel C.,~2005, A\&A 440, 981
\bibitem{} Talon S., Kumar P., Zahn J.-P.,~2002, ApJL 574, 175
\bibitem{} Talon S., Richard O., Michaud D.,~2006, ApJ 645, 634
\bibitem{} Talon S., Vincent A., Michaud G., Richer J.~2003, J. Comp. Phys. 184,
244
\bibitem{} Talon S., Zahn J.-P.,~1997, A\&A  317, 749
\bibitem{} Talon S., Zahn J.-P., Maeder A., Meynet G.,~1997, A\&A  322, 209
\bibitem{} Tassoul J.-L, Tassoul M.,~1982, ApJs 49, 317
\bibitem{} Tassoul M., Tassoul J.-L.,~1983, ApJ  271, 315
\bibitem{} Tayler R.J.,~1973, MNRAS 161, 365
\bibitem{} Taylor G.I.~1923, Phil. Trans. Roy. Soc. A223, 289
\bibitem{} Toqu\'e N., Ligni\`eres F., Vincent A., 2006, GAFD 100, 85
\bibitem{} Townsend A.A.,~1965, J. Fluid Mech. 22, 241
\bibitem{} Townsend R.H.D.,~2003, MNRAS 340, 1020
\bibitem{} Turcotte S., Richer J., Michaud G., Iglesias C.A., Rogers 
F.J.,~1998, ApJ 504, 539
\bibitem{} Unno W., Osaki Y., Ando H., Saio H., Shibahashi H.,~1989, ``Nonradial
oscillations of stars'', $2^{\rm nd}$ edition, University of Tokyo Press
\bibitem{} Urpin V.A., Shalybkov D.A., Spruit H.C.,~1996, A\&A 306, 455
\bibitem{} VandenBerg D.A., Richard O., Michaud G., Richer J.,~2002, ApJ 571,
487
\bibitem{} Vauclair S.,~1983, Astrophysical Processes in Upper Main Sequence
Stars, 13$^{\rm th}$ SAAS-FEE course, Eds. B. Hauck \& A. Maeder
\bibitem{} Vincent A., Michaud G., Meneguzzi M.,~1996, Phys. Fluids 8 (5) 1312
\bibitem{} Vogt H.,~1925, Astron. Nachr. 223, 229
\bibitem{} von Zeipel H.,~1924, MNRAS 84, 665
\bibitem{} Weber E.J., Davis L.,~1967, ApJ 148, 217
\bibitem{} Wendt F.,~1933, Ing. Arch. 4, 577
\bibitem{} Young P.A., Knierman K.A., Rigby J.R., Arnett D.,~2003, ApJ 595,
1114
\bibitem{} Zahn J.-P.,~1975, A\&A 41, 329
\bibitem{} Zahn J.-P.~1983, Astrophysical Processes in Upper Main Sequence
Stars, 13$^{\rm th}$ SAAS-FEE course, Eds. B. Hauck \& A. Maeder
\bibitem{} Zahn J.-P.,~1992, A\&A 265, 115
\bibitem{} Zahn J.-P., Mathis S., Brun A.S.,~2007, A\&A submitted
\bibitem{} Zahn J.-P., Talon S., Matias J.,~1997, A\&A 322, 320
\end{thebibliography}
\end{document}